\documentclass[useAMS,usenatbib]{mn2e}
 
\usepackage{amsmath}
\usepackage{float}
\usepackage{graphicx}
\usepackage{txfonts}
\usepackage{indentfirst}
\usepackage{color}
\usepackage{ulem}
\usepackage{hyperref}

\newcommand{\eq}[1]{\begin{equation}  #1 \end{equation}}
\newcommand{\eqs}[1]{\begin{equation} \begin{split} #1 \end{split} \end{equation}}

\newcommand{\items}[1]{\begin{itemize} #1 \end{itemize}}
\newcommand{\br}[1]{\left( #1 \right)}

\newcommand{\bb}[1]{\left[ #1 \right]}
\newcommand{\ba}[1]{\left\langle #1 \right\rangle}

\newcommand{\dd}{{\rm d}}

\DeclareMathOperator{\arcsinh}{arcsinh}

\def\apj{ApJ}

\def\apjs{ApJS}

\def\mnras{MNRAS}

\def\prd{Phys.~Rev.~D}

\title[Outer profile of dark matter halos]{The outer profile of dark matter
halos: an analytical approach}

\author[Xun Shi]{Xun Shi \thanks{E-mail:
xun@mpa-garching.mpg.de} \\
Max-Planck-Institut f\"ur Astrophysik,
Karl-Schwarzschild-Stra{\ss}e 1, D-85740 Garching bei M\"unchen, Germany\\
}

\begin{document}


\maketitle
  
\label{firstpage}

\begin{abstract}
A steepening feature in the outer density profiles of dark matter halos
indicating the splashback radius has drawn much attention recently. 
Possible observational detections have even been made for galaxy clusters.
Theoretically, Adhikari et al. have estimated the location of the
splashback radius by computing the secondary infall trajectory of a dark matter shell
through a growing dark matter halo with an NFW profile. However, since they
imposed a shape of the halo profile rather than computing it consistently from
the trajectories of the dark matter shells, they could not provide the full
shape of the dark matter profile around the splashback radius. We improve on
this by extending the self-similar spherical collapse model of Fillmore \&
Goldreich to a $\Lambda$CDM universe. This allows us to compute the dark
matter halo profile and the trajectories simultaneously from the mass accretion
history. Our results on the splashback location agree qualitatively with
Adhikari et al. but with small quantitative differences at large mass accretion
rates. We present new fitting formulae for the splashback radius $R_{\rm
sp}$ in various forms, including the ratios of $R_{\rm sp} / R_{\rm 200c}$ and
$R_{\rm sp} / R_{\rm 200m}$. Numerical simulations have made the puzzling 
discovery that the splashback radius scales well with $R_{\rm 200m}$ but not
with $R_{\rm 200c}$. We trace the origin of this to be the correlated
increase of $\Omega_{\rm m}$ and the average halo mass accretion rate with an
increasing redshift.

\end{abstract}


\section[]{Introduction}

Recent numerical simulations \citep{diemer14} have noticed a sharp
steepening in the outer density profiles of dark matter halos, which seems to
offer a physical boundary of dark matter halos \citep{more15}. Although
this steepening feature lies in a low density region in the outskirts of a dark
matter halo and thus hard to detect, potential observational evidences have been
found by studying the projected number density profiles of galaxies around galaxy
clusters \citep{patej15b, more16}.

Physically, this steepening feature has been identified with the `splashback' of recently
accreted dark matter i.e. piling up of dark matter near the first apocenter of
its orbit through the halo \citep{diemer14}. With this physical picture,
\citet{adhi14} theoretically estimated the radial position of the splashback $R_{\rm sp}$. They
computed $R_{\rm sp}$ by tracing the trajectory of a dark matter shell accreted
onto a dark matter halo and then going through it, and described the dark matter
halo with an NFW profile \citep{nfw96,nfw97} with its mass increasing
with time. They found good agreement of the predicted splashback radius and the
steepening position from the stacked halos in the \citet{diemer14} simulations.  
However, one weakness of their method is that the imposed NFW shape of the
halo profile and the trajectories of the dark matter shells are not fully
consistent with each other. Although this may not affect their estimation for
$R_{\rm sp}$ to a large extent, it prevents them from giving the full shape of
the dark matter profile around the splashback radius.

In fact, the steepening feature at halo outskirts has been known for long time
in the self-similar spherical collapse model \citep{fillmore84, bert85,
lithwick11} as associated with the outmost caustic. In contrast, the caustics in
dark matter halos had not been noticed in three-dimensional numerical
simulations due to resolution limits until lately when special techniques were applied
\citep{vogel09b,hahn16}. 
As the outer profiles of dark matter halos
are primarily determined by the recent mass accretion, and the matter there has
relatively little participation in the relaxation processes which require
numerical simulations to capture, the self-similar spherical collapse model is
actually rather adequate in describing the outer profile of dark matter halos. 
Another merit of the self-similar spherical collapse model is that the halo
profile and the trajectories of the dark matter shells are treated consistently
with iterations of computing one from the other. 

Thus we use the self-similar spherical collapse model to compute the outer
profiles of dark matter halos. One obstacle remains still, that the
self-similarity is strictly valid only in a universe with no characteristic
scales, e.g. an Einstein de-Sitter (EdS) universe ($\Omega_{\rm m}=1$) as
considered in previous studies \citep{fillmore84, bert85, lithwick11}. Here we extend the model to a
$\Lambda$CDM universe by relaxing the self-similarity ansatz while keeping its
feature of computing trajectories and halo profile in a consistent manner. 

The rest of the paper is organised as follows. In Sect.\;\ref{sec:numerical} we
describe the detailed procedures of our computation and the resulting mass and
density profiles of dark matter halos. Then we derive analytical interpretations of
the results, and provide fitting formulae in Sect.\;\ref{sec:analytical},
summarise our findings and conclude in Sec.\;\ref{sec:conclusion}.
We present the minimalistic dimensionless form of spherical collapse in a
$\Lambda$CDM universe in Appendix.\;\ref{app:spherical_collapse}, which
scales out the cosmology dependence in the variables and has thus greatly
reduced our computational cost. With this form, many relations among the
variables as well as the dark matter density profile in the single stream
regime of spherical collapse can be expressed in simple analytical forms, which we show in
Appendices.\;\ref{app:eqs} and \ref{app:single}.
Finally, in Appendix.\;\ref{app:splashback_EdS} we present a physical
estimation of the splashback radius in an EdS universe.

\section[]{Mass and density profiles from profile-trajectory iteration}
\label{sec:numerical}

\subsection[]{Methods}
We look for the asymptotic profile of dark matter halos as a function of their mass accretion rate and the background cosmology.
For simplicity, we restrict ourselves to a flat $\Lambda$CDM
universe with $\Omega_{\rm m} + \Omega_{\rm \Lambda} = 1$, and study the case of
power law mass growth $M_{\rm vir} \propto a^s$ where $a$ is the cosmic scale
factor. Thus, the two basic parameters in our study are the mass accretion rate $s$ and the dimensionless
matter density of the universe $\Omega_{\rm m}(a)$.

Power law mass growth is typical for dark matter halos in our Universe
at low redshifts \citep[e.g.][]{correa15}. The typical mass
growth rate is $s \lesssim 1$ for galaxy mass halos today, and increases with redshift and halo
mass. Following \citet{adhi14}, we use the mass
enclosed by the shell that has collapsed to half of its turn-around radius
$M_{\rm hrta}$ under no shell-crossing as a proxy for
the virial mass $M_{\rm vir}$ of the halo. This matches the
early definition of virial radius in a spherical collapse model \citep{lacey93}.
With this definition, the relative sizes of dark matter shells are specified by
their dimensionless trajectories (Appendix.\;\ref{app:spherical_collapse}) when
the mass growth history is given.

\begin{figure}
\centering
    \includegraphics[width=.45\textwidth]{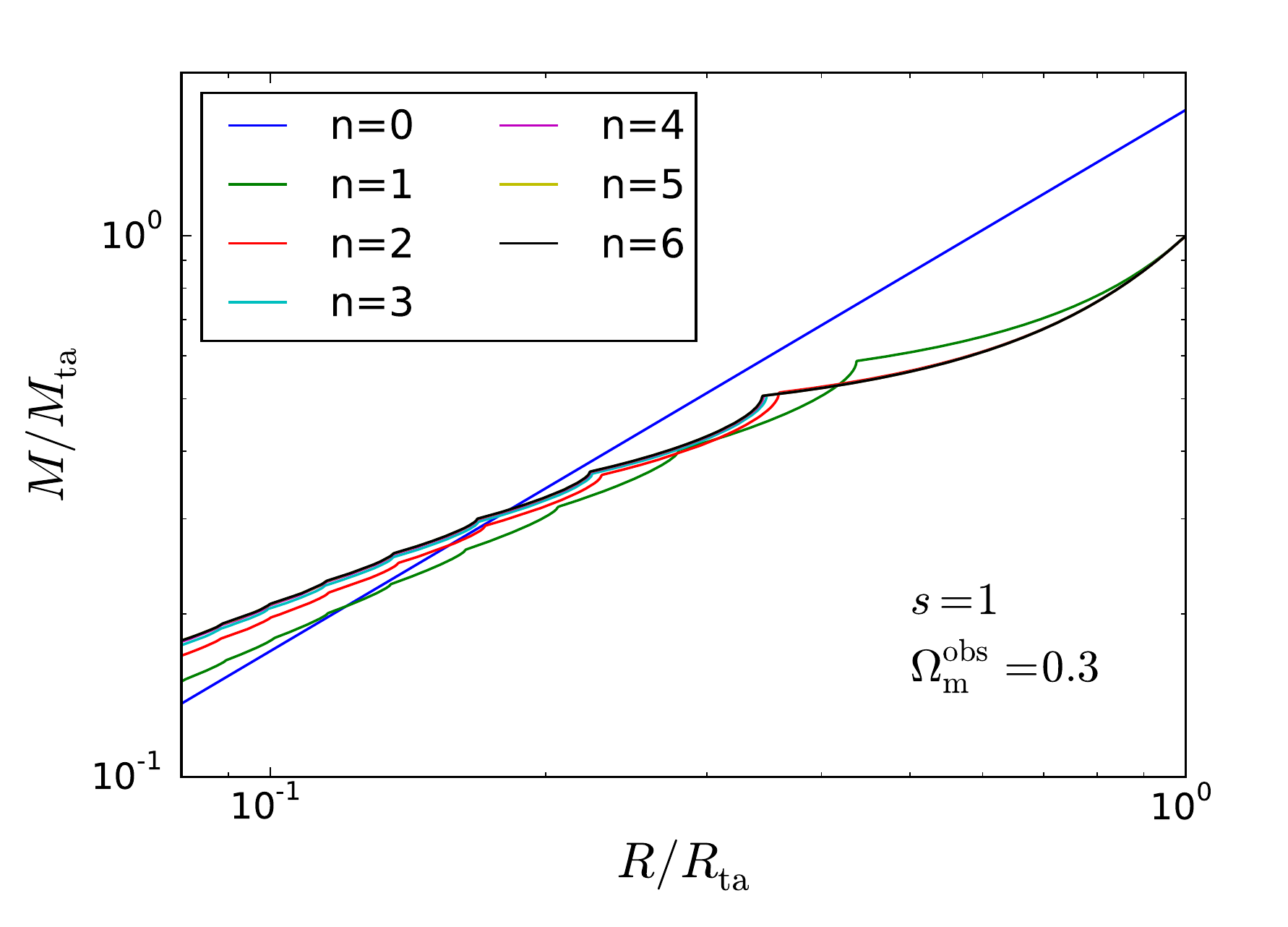} 
  \caption{Mass profile converges quickly with the number of iteration $n$.
  The plotted mass is scaled to $M_{\rm ta}$, the mass within the
  turn-around radius $R_{\rm ta}$.}
\label{fig:iter}
\end{figure} 
 
\begin{figure}
\centering
    \includegraphics[width=.45\textwidth]{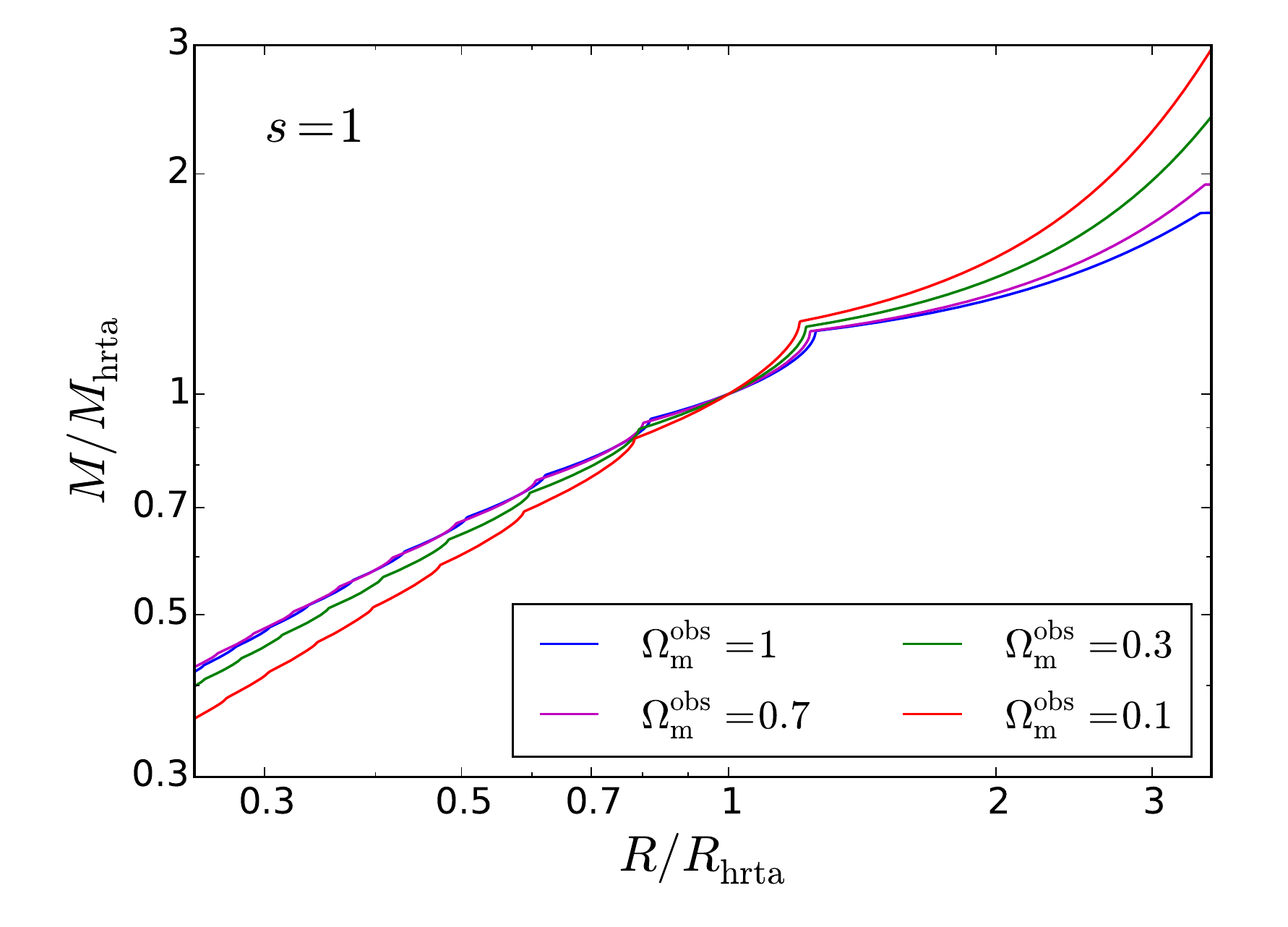}
  \caption{Dependence of the halo mass profile on $\Omega_{\rm m}^{\rm obs}$.}
\label{fig:mprof_omm}
\end{figure}

In a multi-stream region such as a dark matter halo, the dark matter trajectory
depends on the halo mass profile and vice versa, thus an iteration is needed in 
computing the two. Under strict self-similarity, the trajectory of one dark
matter shell also represents the locations of all dark matter shells in a
snapshot, and thus within each iteration, only one integration of orbit is
needed.
In a $\Lambda$CDM universe, we lose such simplification, and need to compute
the trajectories of the dark matter shells accreted at different times
simultaneously. To reduce this computational cost, we simplify the spherical
collapse equations to a minimalistic dimensionless form
(Appendix.\;\ref{app:spherical_collapse}). The loss of strict self-similarity
also means that the mass profile is time dependent. Luckily, in our Universe
where the dark energy is still not entirely dominating, this time dependence is
mild and slow - on the timescale of change of
$\Omega_{\rm m}(a)$, see Fig.\;\ref{fig:mprof_omm}. Over a 
dynamical time which is shorter compared to this timescale, the mass profile can
be approximated as being constant.
 
Therefore, we parametrize the mass in the multi-stream region as 
$M(R) = M_{\rm hrta}(a) f(R/R_{\rm hrta})$, with an initial choice of $f_0(x) =
x$. Then we compute the trajectories of the shells till the time the halo is
observed, which is specified by the value of $\Omega_{\rm m}^{\rm obs}$. Using
the trajectories we update the form of mass profile to $f_1(x)$, compute the
trajectories again, and iterate. The mass profile and trajectories converge
quickly with the number of iterations $n$ (Fig.\;\ref{fig:iter}). As shown by
Fig.\;\ref{fig:mprof_omm}, the shape of the mass profile scaled to $R_{\rm
hrta}$ is indeed only slightly time-dependent, even over a long time range
indicated by the change of the matter content of the universe $\Omega_{\rm
m}^{\rm obs}$. 

\subsection[]{Results}

\begin{figure*}
\centering
  \begin{tabular}{@{}c}
    \includegraphics[width=0.9\textwidth]{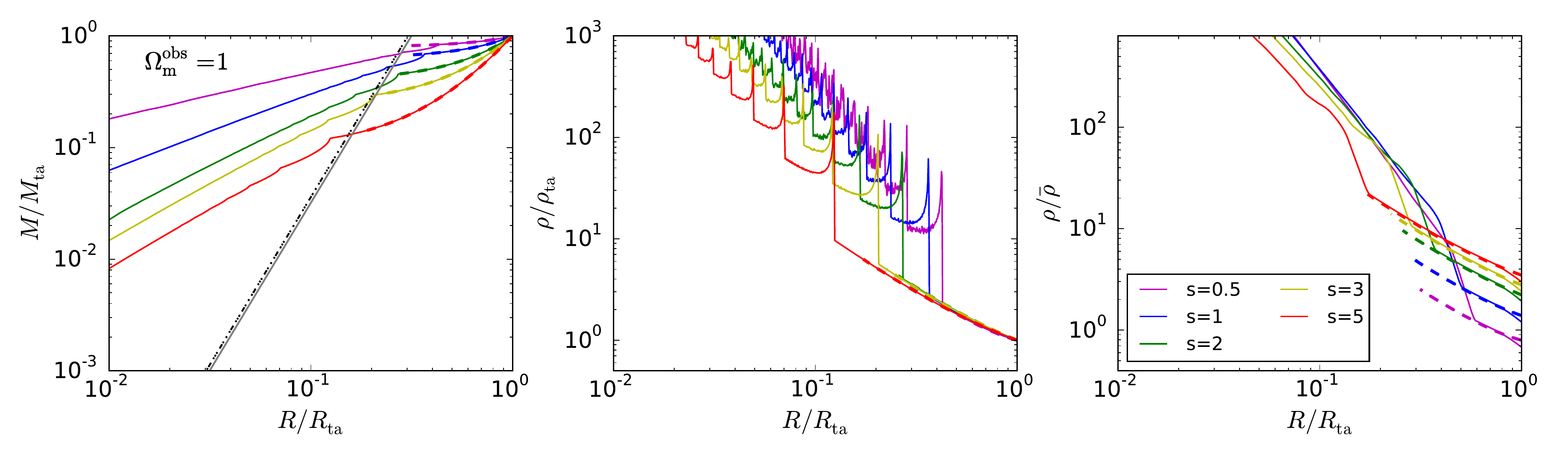}\\   
    \includegraphics[width=0.9\textwidth]{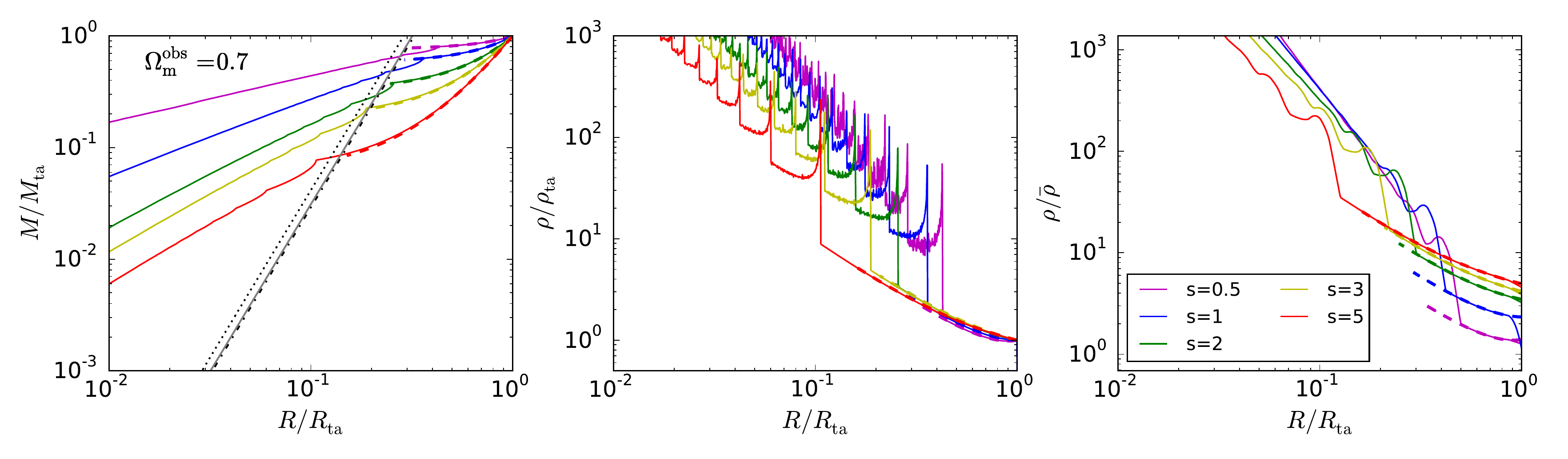}\\
    \includegraphics[width=0.9\textwidth]{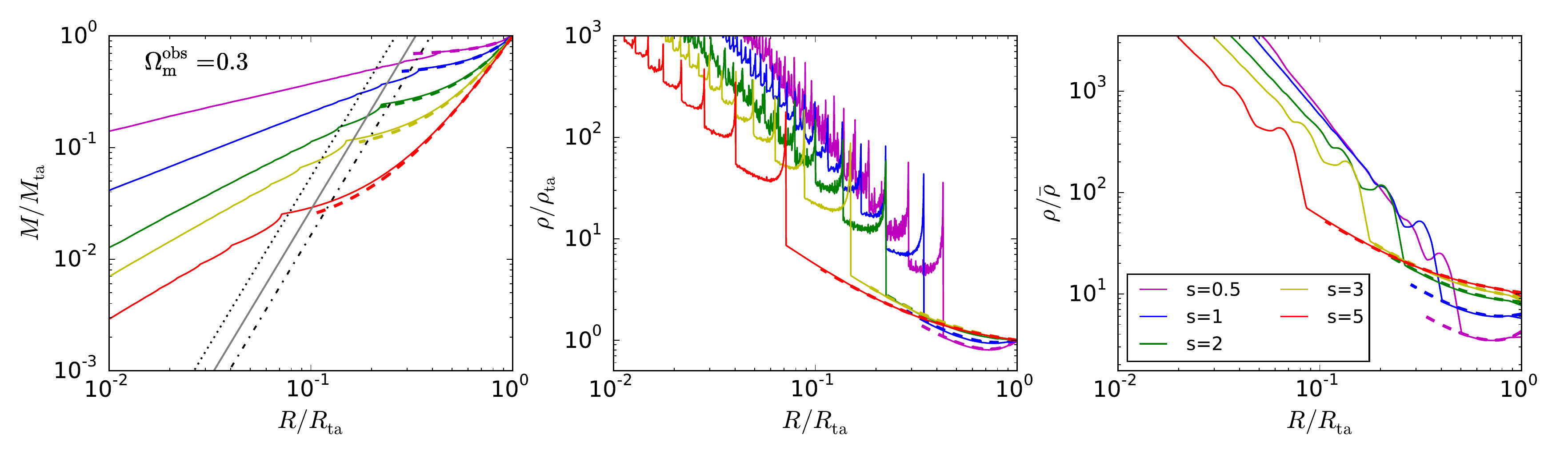}\\   
    \includegraphics[width=0.9\textwidth]{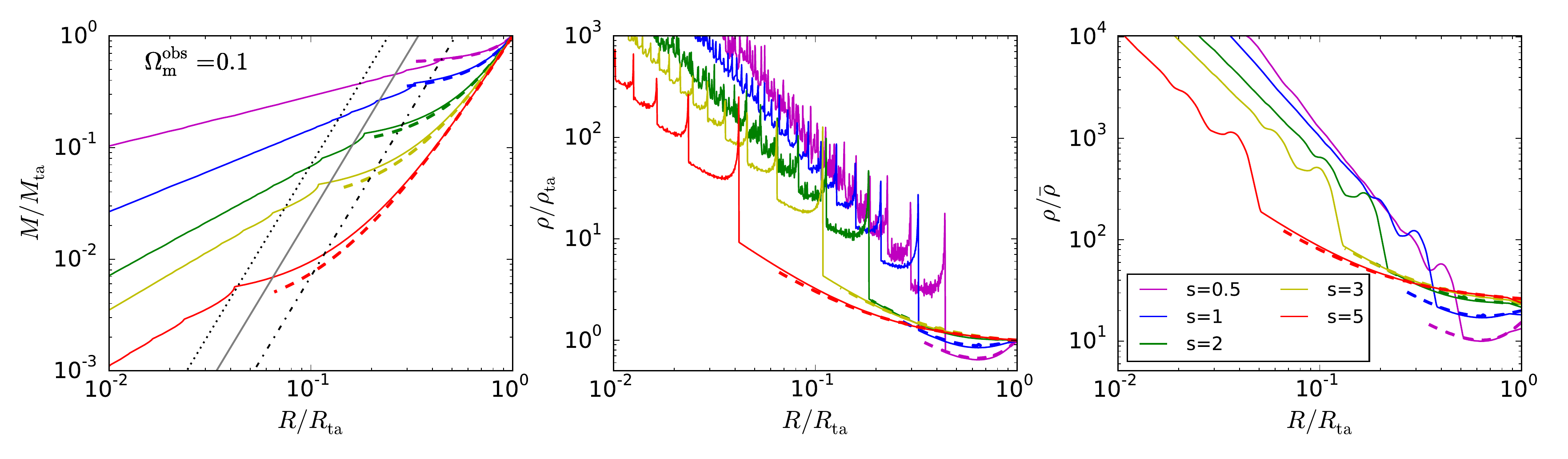} 
  \end{tabular}
\caption{Profiles of mass (left), density (middle) and smoothed density
(right) for different values of $\Omega_{\rm m}^{\rm obs}$ (different rows)
and different mass accretion rate $s$ (colored lines in each panel) from the
mass profile - trajectory iterations. The color dashed lines show the
analytical mass and density profile given by Eq.\;(\ref{eq:accdens}). The
black dotted and dashed lines in the left column mark the
locations of $\Delta_{\rm c} = 200$ and $\Delta_{\rm m} = 200$ respectively,
and intersection between the gray solid line and the color lines show the
locations of the virial radius.}
\label{fig:all}
\end{figure*}

\begin{figure}
\centering
    \includegraphics[width=.47\textwidth]{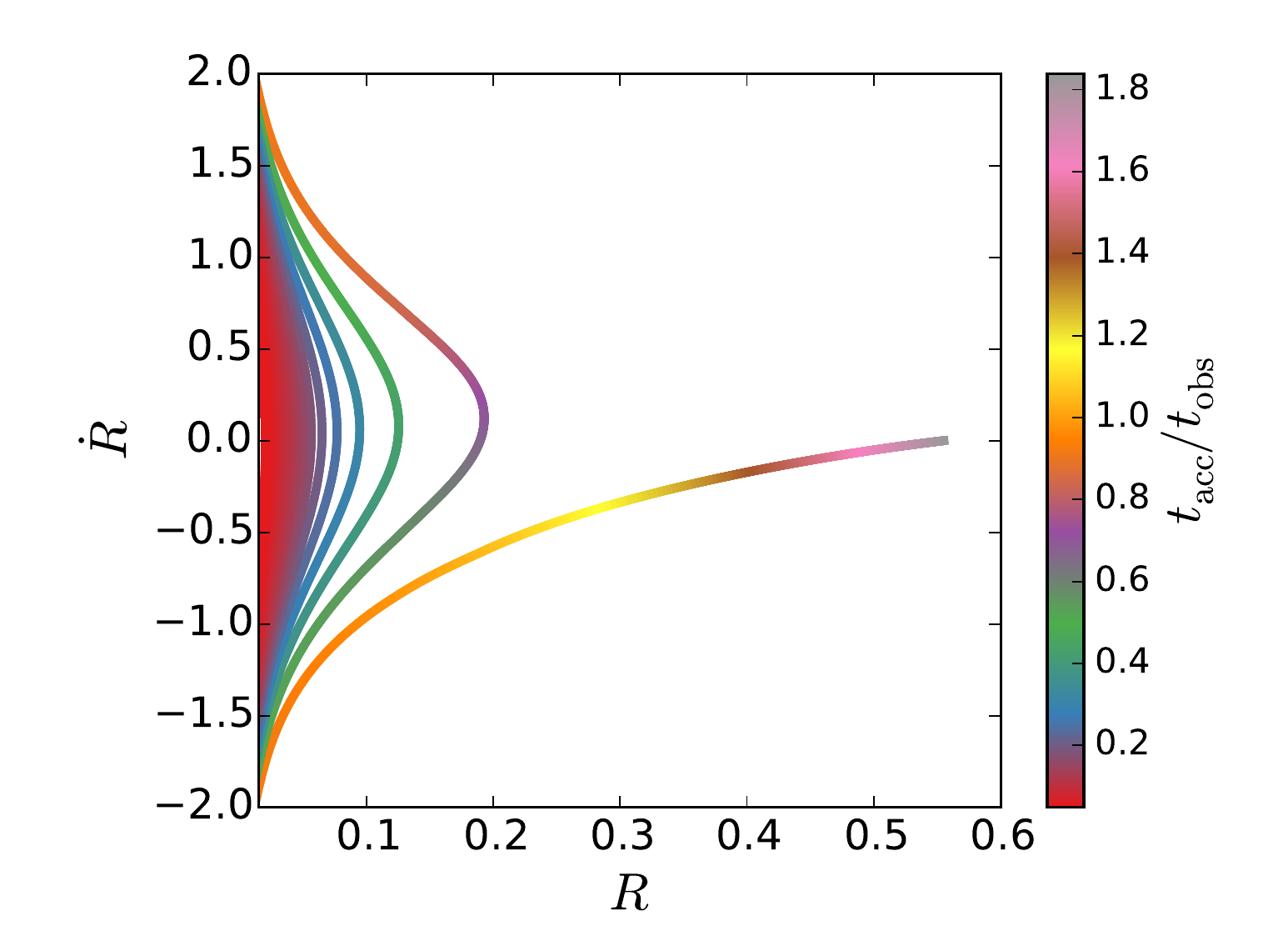}
  \caption{Instantaneous phase space locations of the shells at the time of
  observation when $\Omega_{\rm m}^{\rm obs} = 0.3$. The color coding shows the
  time when the shells are accreted by the halo $t_{\rm acc}$ compared to the
  time of observation $t_{\rm obs}$.
  Here, the accretion rate $s=1$.} 
\label{fig:phase}
\end{figure}

The converged mass profiles, the density profiles derived from them, and their
dependencies on $\Omega_{\rm m}^{\rm obs}$ and the mass accretion rate $s$ are
shown in Fig.\;\ref{fig:all}. We find three features that characterise the
general shapes of the mass and density profiles: a power-law inner slope, the
splashback feature, and a smooth profile outside the splashback radius where the
matter is still on its infall towards the halo.
The other caustics in the density profiles (middle column of Fig.\;\ref{fig:all}) are hardly observable
after smoothing (right column of Fig.\;\ref{fig:all}) which in reality would be
caused by e.g. asphericity of the halo and
instability to perturbations \citep{hen99}.

The power-law inner mass profile $M \propto R^{\Upsilon}$ has been studied by
\citet{fillmore84} in an EdS universe. There, the mass accretion rate is related
to the logarithmic slope $\epsilon$ of the initial mass perturbation $\delta
M_i / M_i \propto M_i^{-\epsilon}$ with $s = 1/\epsilon$.
In terms of $s$, the \citet{fillmore84} result reads
\eq{
 \label{eq:gamval}
  \Upsilon =
  \begin{cases}
    3s/(s+3)  & \quad \text{if } s \le 3/2\\
    1       & \quad \text{if } s \ge 3/2 \,.\\
  \end{cases}
}
We reproduce this inner slope dependence on $s$ not only when $\Omega_{\rm
m}^{\rm obs}=1$, but also for other $\Omega_{\rm m}^{\rm obs}$ values (first
column of Fig.\;\ref{fig:all}). This is expected since the inner profiles are
composed primarily of shells collapsed at early times when dark matter
dominates the energy content of the universe. Admittedly, in the very
inner regions of dark matter halos, especially within the scale radius of an NFW
profile, the dynamics of dark matter particles is heavily influenced by
relaxation mechanisms including the radial orbit instability
\citep[e.g.][]{merritt85,macmillan06,vogel11b}, and the angular momentum
of the particles cannot be neglected. The one-dimensional accretion model we use naturally cannot reproduce the true
dark matter profiles in these regions. However, as the dynamical time in the
central region is very short, the inner profile should not have much
influence on the dark matter dynamics and profiles in the outer
regions.

The sharp jump in the density profile at the splashback radius is evident in all
cases, and the amplitude of the density jump for the un-smoothed density profile
always lies around a factor of $4$ to $5$. This can be understood considering
that the splashback radius separates the regions with one and three dark matter
streams (Fig.\;\ref{fig:phase}), and that the earlier accreted matter is denser
than the matter currently being accreted. 
What depends on the accretion rate and the cosmology is the location of the
splashback radius, as has been discovered by previous studies
\citep{adhi14,diemer14}. As shown by Fig.\;\ref{fig:all}, the ratio of the
splashback radius and the turn-around radius decreases with the
accretion rate $s$. For low accretion rates, the ratio is approximately
independent of $\Omega_{\rm m}^{\rm obs}$, whereas the decreasing with $s$ at
high accretion rate is more significant for a universe with low matter content.
We investigate the locations of the splashback radius in more details in
Sect.\;\ref{sec:ana_sp}, where we provide explanations to these behaviors.
 
Outside the splashback radius and within the turn-around radius, the matter
is still on its initial infall onto the halo. The dark matter density profiles
in this regime (color solid lines in the middle and right
columns of Fig.\;\ref{fig:all} outside the splashback radius) represent the
correlated matter around the halo, or the so-called 2-halo term. Instead of
modeling the profiles in this region in an averaged sense from the matter
power spectrum, we can describe them using the spherical collapse model to take
account of the dependence on the accretion rate of the individual halos. Indeed,
the numerically integrated profiles match well with the single-stream
analytical profiles (color dashed lines) given in Appendix.\;\ref{app:single}.

\section[]{Analytical interpretations and fitting formula}
\label{sec:analytical}

\subsection[]{The splashback radius}
\label{sec:ana_sp}

\begin{figure}
\centering
    \includegraphics[width=.4\textwidth]{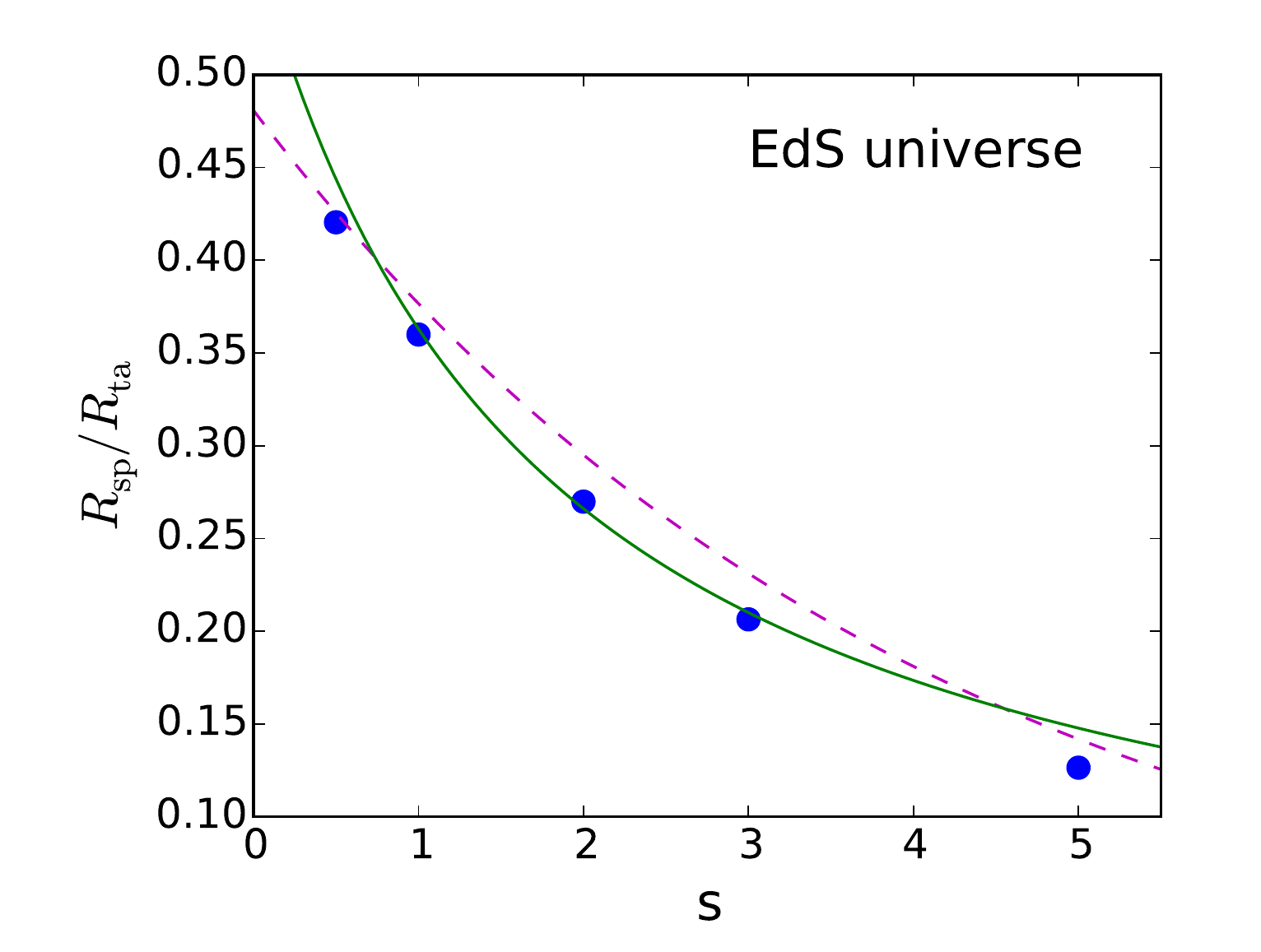} 
  \caption{The ratio of splashback radius and the turnaround radius at the
  time of observation in an EdS universe as a function of accretion rate $s$.
  The analytical formula given in Eq.\;(\ref{eq:Rsp2ta_all}) for $s \le 3/2$ (magenta dash
  line) and $s \ge 3/2$ (green solid line) agree well with
  the numerical results (blue circles).}
\label{fig:rsp}
\end{figure}

After turning around, dark matter shells begin to oscillate in the gravitational
potential of the dark matter halo. 
Approximating the mass profiles by their inner power law
shape, $M = (R/R_{\rm ta})^{\Upsilon} M_{\rm ta}$ with $\Upsilon$ given by
Eq.\;(\ref{eq:gamval}), one can see that the gravitational
potential of the dark matter halo in a Eulerian coordinates grows with time
when the accretion rate $s > 3/2$ while stays approximately constant when $s \le
3/2$. 

We study the location of the splashback radius by tracing the dynamics of the
oscillating orbits in these two regimes. In an EdS universe, approximate
analytical solutions to the dynamical equation can be obtained as
(Appendix.\;\ref{app:splashback_EdS})
\eqs{
\label{eq:Rsp2ta_all}
\frac{R_{\rm sp}}{R_{\rm ta}}  \approx  
  \begin{cases}
3^{-2/3 - 2s/9}  & \quad \text{when } s \le 3/2 \,,\\
\bb{1 + 4(4s/9+1/3)/\sqrt{\pi}}^{-1} & \quad \text{when } s \ge 3/2 \,,
  \end{cases}
}
which well-describes the smaller ${R_{\rm sp}}/R_{\rm ta}$ ratio for the
range of mass accretion rates we have tested (Fig.\;\ref{fig:rsp}).
Different approximations are taken for the two regimes $s \le 3/2$ and $s
\ge 3/2$ and they lead to slightly different ${R_{\rm sp}}/R_{\rm ta}$ values at
$s = 3/2$ (see Appendix.\;\ref{app:splashback_EdS} for details).

In a $\Lambda$CDM universe, the growth of the turn-around radius with time is
more significant at a certain mass accretion rate compared to the case of an
EdS universe, due to the additional expansion caused by the dark energy. This
explains the smaller ${R_{\rm sp}}/R_{\rm ta}$ ratio when $\Omega_{\rm m}^{\rm
obs}$ is smaller (Fig.\;\ref{fig:all}). This effect is more pronounced for a
larger mass accretion rate. In the limit of small accretion rate $s \to 0$, the
${R_{\rm sp}}/R_{\rm ta}$ ratio stays approximately invariant with the
change of the cosmological background. As evident in Fig.\;\ref{fig:all}, for
the lowest accretion rate we sampled, $s=0.5$, the position of the splashback
radius already approaches the asymptotic value of $R_{\rm sp} / R_{\rm ta} =
3^{-\frac{2}{3}} \lesssim 0.5$.

\subsection[]{Time when the splashback matter was accreted} 
\label{sec:time}

\begin{figure}
\centering
    \includegraphics[width=.47\textwidth]{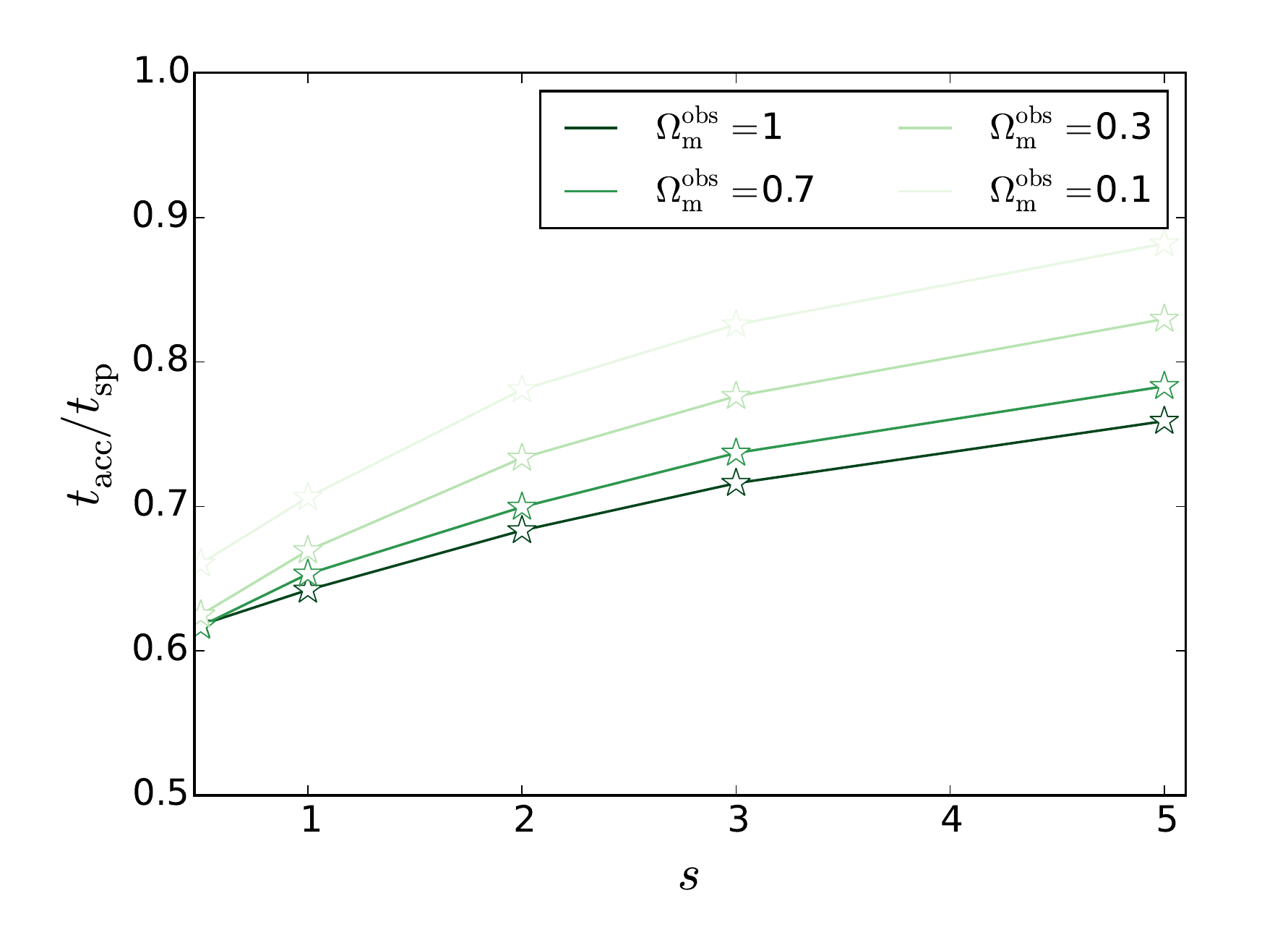} 
  \caption{The time when the matter at the splashback radius was accreted
  $t_{\rm acc}$ compared to the time it is observed at the splashback radius
  $t_{\rm sp}$.}
\label{fig:tacc}
\end{figure}
Since the actual mass accretion rate of a dark matter halo changes with
time, how to match the mass accretion rates in simulations to the
idealised accretion rate $s$ defined from a power-law mass accretion
history is non-trivial.
Current numerical studies commonly apply $\Gamma = \Delta \log M / \Delta \log
a$ over a fixed cosmic scale factor interval to the simulated halos \citep[e.g.][]{diemer14,adhi14,lau15},
and $\Gamma \approx s$ is often assumed when comparing to analytical results.
This simplified assumption is possibly the biggest limitation to the precision
of the comparison between the analytical predictions and the numerical results. To improve on this, it is useful to know the time
when the splashback matter was accreted. 

Regarding the orbit of a dark matter shell as an oscillation starting from the
beginning of its expansion, the turn-around happens at phase $1/4$ of the
first oscillation, accretion onto the halo happens at around phase $1/2$, and
splashback happens at phase $3/4$.
Thus, the age of the universe at which the matter currently at the splashback radius was
accreted on to the halo $t_{\rm acc}$\footnote{Here, we use $t_{
\rm hrta}$, the time when the shell collapses to half its turn-around radius, to
define the `accretion time' $t_{\rm acc}$.} should be close to $(1/2)/(3/4) =
2/3$ of the current age $t_{\rm sp}$ at low accretion rates where the halo potential is roughly constant with time. This rough estimate is supported by the results from numerical integration
shown in Figs.\;\ref{fig:phase} and \ref{fig:tacc}. In detail, the ratio of
$t_{\rm acc}/t_{\rm sp}$ increases with increasing accretion rate and decreasing
matter density of the universe. This means, for example, that the splashback
matter at redshift zero in a flat universe with $\Omega_{\rm m0}=0.3$ was
accreted at approximately redshift $z \approx 0.4$ for an accretion rate $s=1$
and $z \approx 0.2$ for an accretion rate $s=5$. This gives a hint on how to
choose the redshift range to average the mass accretion rate measured in
simulations for a comparison study.

\subsection[]{Fitting formulae and comparison to previous studies} 
\label{sec:fit}
\begin{figure}
\centering
  \begin{tabular}{@{}c}
    \includegraphics[width=0.45\textwidth]{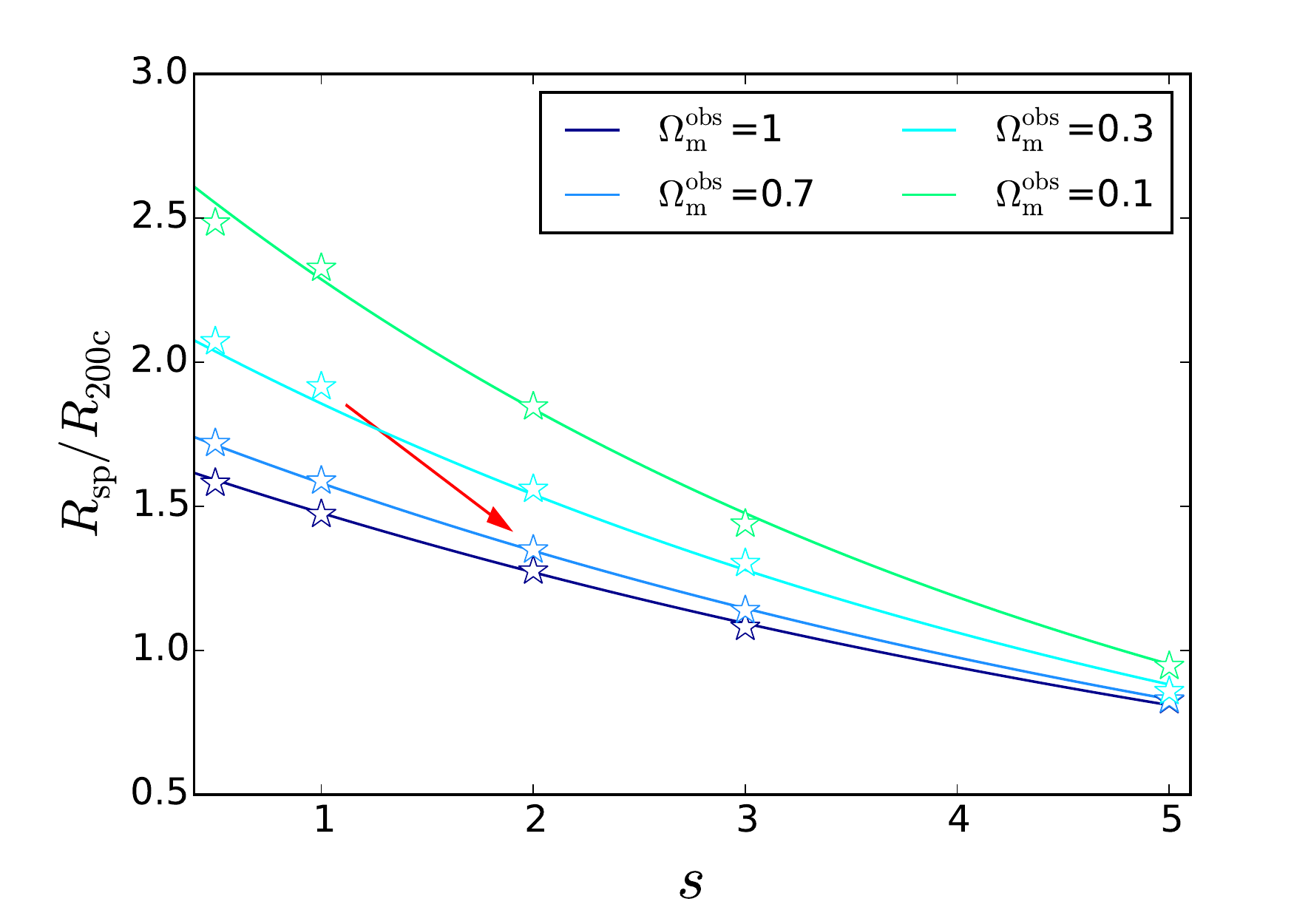}\\   
    \includegraphics[width=0.45\textwidth]{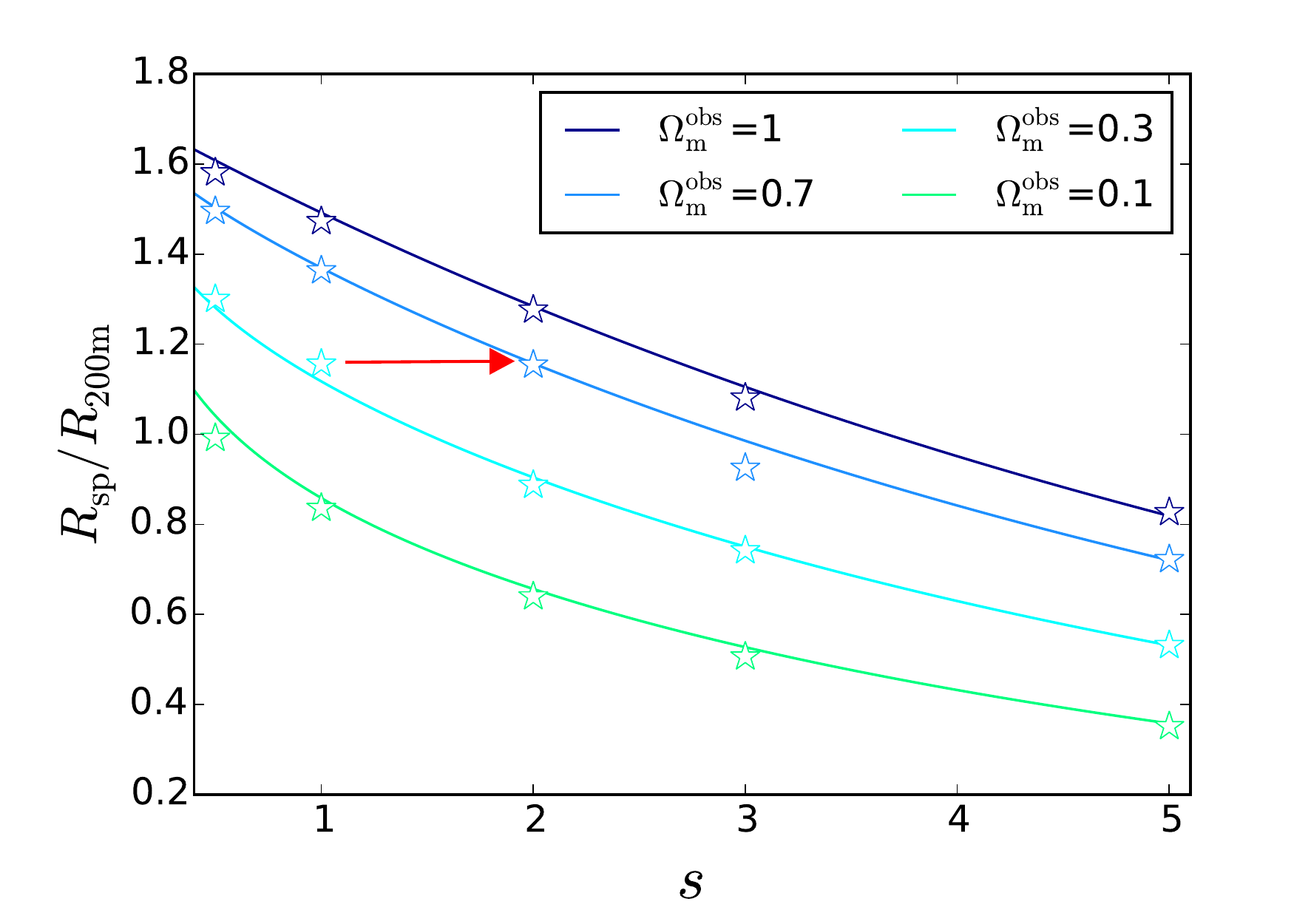}\\
    \includegraphics[width=0.45\textwidth]{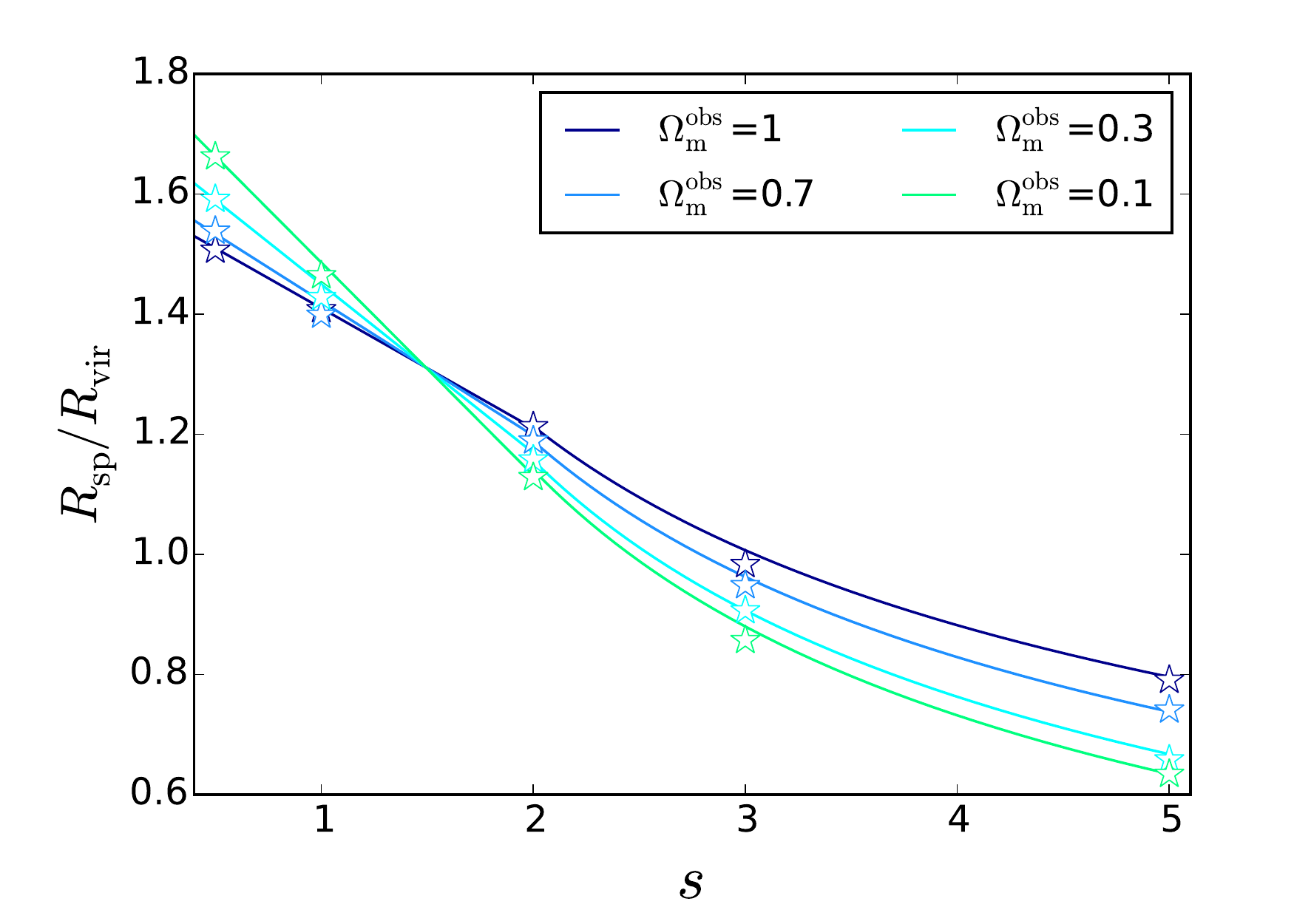}
  \end{tabular}
\caption{The location of the splashback radius with respect to $R_{\rm 200c}$
(upper panel), $R_{\rm 200m}$ (middle panel) and $R_{\rm vir}$ (lower panel).
Values from numerical integration are shown as the stars. The lines are from
fitting formulae Eqs.\;\ref{eq:fit200c}, \ref{eq:fit200m} and \ref{eq:fitvir},
respectively. The red arrows in the upper and middle panels demonstrate how
the $R_{\rm sp}/R_{\rm 200c}$ and $R_{\rm sp}/R_{\rm 200m}$ ratios would
typically change between halo samples observed at two redshifts. That the
orientation of the arrow for $R_{\rm sp}/R_{\rm 200m}$ would typically be more
horizontal explains why $R_{\rm sp}$ scales better with $R_{\rm 200m}$ than
$R_{\rm 200c}$ (see text for more details).}
\label{fig:fitting}
\end{figure}

\begin{figure}
\centering
    \includegraphics[width=.47\textwidth]{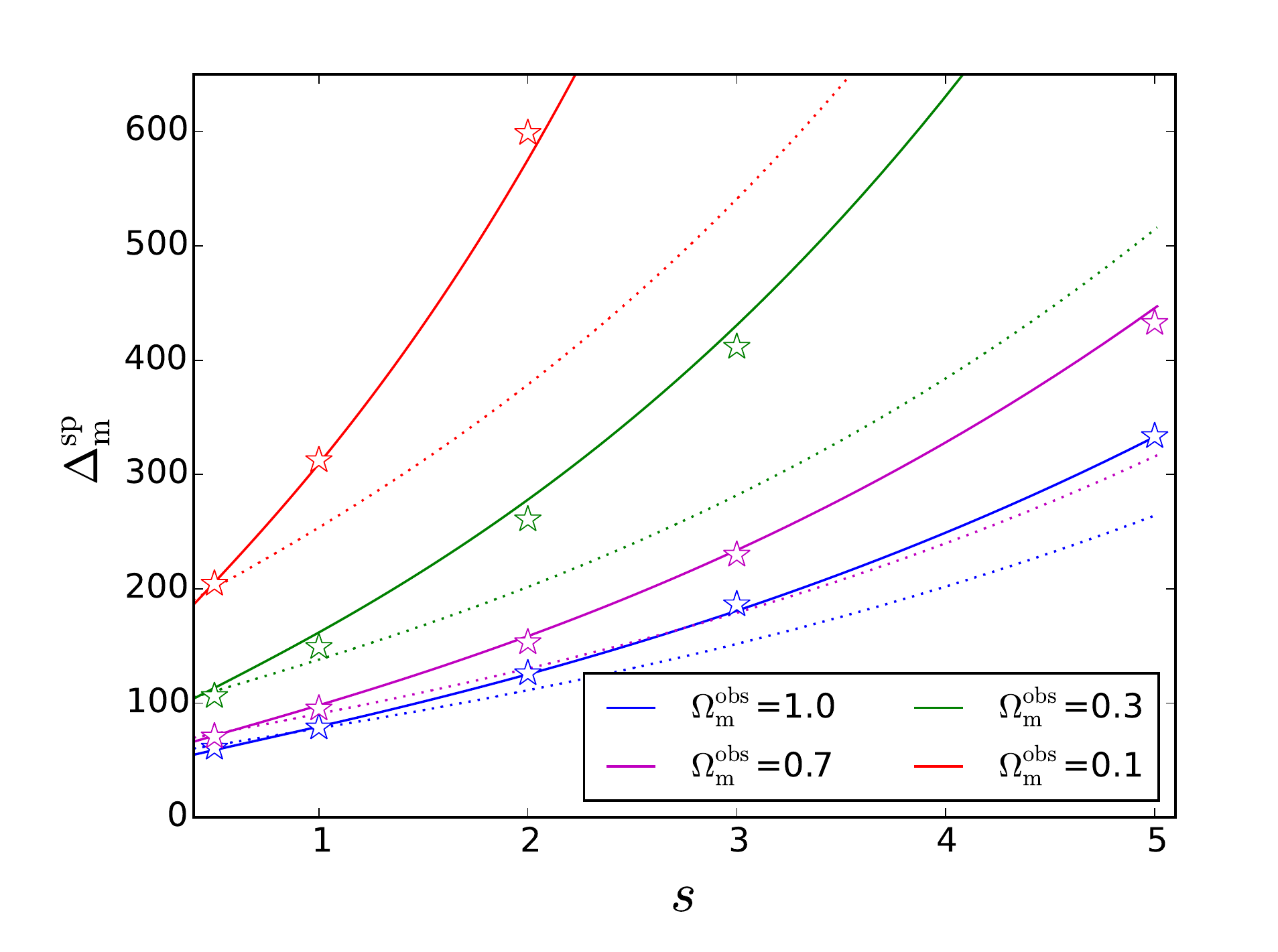} 
  \caption{The overdensity at the splashback radius. Solid lines show fitting
  formula Eq.\;\ref{eq:fitdelta} to our results from numerical integration
  (stars). In comparison, the \citet{adhi14} fitting formula is shown as the
  dotted lines.}
\label{fig:delsp}
\end{figure}

Since we have computed the halo mass profile and the splashback location
consistently, we can study the splashback location with respect to the
commonly used radii defined from the halo mass profile, e.g. $R_{\rm 200c}$,
$R_{\rm 200m}$ and $R_{\rm vir}$.  
Here we provide fitting formulae for the locations of the splashback radius in
terms of these radii as functions of the mass accretion rate $s$ and the matter
content of the universe $\Omega_{\rm m}^{\rm obs}$.

\eq{
\label{eq:fit200c}
\frac{R_{\rm sp}}{R_{\rm 200c}} =
{\rm Exp}\bb{(-0.22+0.03s)\ln\Omega_{\rm m}^{\rm obs}+0.54-0.15s} \,,
}
\eq{
\label{eq:fit200m}
\frac{R_{\rm sp}}{R_{\rm 200m}} =
{\rm Exp}\bb{(0.24+0.074\ln s)\ln\Omega_{\rm m}^{\rm obs}+0.55-0.15s} \,,
}

\eqs{
\label{eq:fitvir}
& \frac{R_{\rm sp}}{R_{\rm vir}} = \\
  &\begin{cases}
     (-0.2+0.067\ln\Omega_{\rm m}^{\rm obs})\ln s + 1.6
    - 0.1\ln\Omega_{\rm m}^{\rm obs}  & \text{if } s
    \le 2 \,;\\
    {\rm Exp}\bb{(-0.66+0.2\ln\Omega_{\rm m}^{\rm obs}) \ln
    s+0.58-0.07\ln\Omega_{\rm m}^{\rm obs}}  
     &\text{if } s > 2 \,.\\
  \end{cases}
}
The locations of $R_{\rm 200c}$, $R_{\rm 200m}$ and $R_{\rm vir}$ are
determined using the mass profiles, and $R_{\rm sp}$ using the outmost radius
where the density diverges in the un-smoothed density
profiles.
The mean overdensity within $R_{\rm vir}$ at virial radius depends on cosmology as \citep{bryan98}
\eq{
\Delta_{\rm c, vir}(\Omega_{\rm m}) = 18\pi^2 + 82(\Omega_{\rm m}-1) -
39(\Omega_{\rm m}-1)^2 
}
\citep[see also][for other fitting formulae]{lacey93,naka97}.

For other reference radii e.g. $R_{\rm 500m}$, one can use $M \propto
R^{\Upsilon}$ with $\Upsilon$ given by Eq.\;(\ref{eq:gamval}) as an approximation of the outer
mass profile to scale the reference radii to $R_{\rm 200c}$ or $R_{\rm 200m}$.
For example, $R_{\rm 500m}/R_{\rm 200m} = (500/200)^{1/(\Upsilon-3)}$. This holds
approximately correct as long as the chosen reference radius is large enough
that relaxation physics is still sub-dominant compared to accretion.

Fig.\;\ref{fig:fitting} shows the numerically integrated results for the
locations of the splashback radius in comparison with the fitting formulae. In
units of $R_{\rm 200c}$ or $R_{\rm 200m}$, the splashback radius also decreases with increasing accretion rate $s$ (Fig.\;\ref{fig:fitting}),
consistent with numerical findings \citep{diemer14,lau15} and analytical
results from \citet{adhi14}. We also notice the good
agreement on the accretion rate dependence of $R_{\rm sp} / R_{\rm 200m}$
between our result and the numerical result of \citet{more15} (their Fig.\;3)
within the tested range of mass accretion rates and redshifts, despite of
the imperfect match between $s$ and the accretion rate indicator measured from simulations. The dependence on $\Omega_{\rm m}^{\rm obs}$ differs
when scaled with $R_{\rm 200c}$ or $R_{\rm 200m}$. The ratio $R_{\rm sp}/R_{\rm
200c}$ decreases with increasing $\Omega_{\rm m}^{\rm obs}$ at small accretion
rate, while saturates to $R_{\rm sp}/R_{\rm 200c} \lesssim 1$ at large accretion rate for all $\Omega_{\rm
m}^{\rm obs}$ values. In comparison, the ratio $R_{\rm
sp}/R_{\rm 200m}$ increases with $\Omega_{\rm m}^{\rm obs}$ at all accretion
rates (Fig.\;\ref{fig:fitting}). 

Numerical studies \citep{diemer14,lau15} have found that the splashback
radius of halos at different redshifts align much better with each other
when scaled with $R_{\rm 200m}$ than $R_{\rm 200c}$. The origin of this
intriguing property has not yet been understood. From our results
(see e.g. Fig.\;\ref{fig:fitting}), the alignment of the splashback radius is
\textit{not} better with $R_{\rm 200m}$ than $R_{\rm 200c}$ at a fixed accretion
rate for the cosmology and redshift range considered in the simulations
($\Omega_{\rm m}^{\rm obs} \approx 0.3$ or greater). The apparent better
alignment when scaled with $R_{\rm 200m}$ is an outcome of correlated
increase of $\Omega_{\rm m}^{\rm obs}$ and the average halo mass accretion rate
with redshift:
the decrease of $R_{\rm sp} / R_{\rm 200m}$ with $s$ and the increase with
$\Omega_{\rm m}^{\rm obs}$ cancel each other to some extent; on the other hand
$R_{\rm sp} / R_{\rm 200c}$ decreases with both $\Omega_{\rm m}^{\rm obs}$ and
$s$, which leads to an inevitable decrease of $R_{\rm sp}/R_{\rm 200c}$ with redshift (see the
red arrows in Fig.\;\ref{fig:fitting}).

Another way of describing the location of the splashback radius is to look
at the mean overdensity within it $\Delta_{\rm m}^{\rm sp}$. However, this
representation is more susceptible to the inner mass profile and thus demands a more careful description of it
taking account of the relaxation processes. Therefore, the self-similar
solutions are not best represented in terms of $\Delta_{\rm m}^{\rm sp}$.
Nevertheless,
we provide a fitting formula to $\Delta_{\rm m}^{\rm sp}$, 
\eq{
\label{eq:fitdelta}
\Delta_{\rm m}^{\rm sp} := \frac{\ba{\rho}(<R_{\rm sp})}{\bar{\rho}}
= 33\br{\Omega_{\rm m}^{\rm obs}}^{-0.45}{\rm
Exp}\bb{\br{0.88-0.14\ln\Omega_{\rm m}^{\rm obs}}s^{0.6}} \,,
} 
and compare it to the previous work of \citet{adhi14}.
Our fitting formula (solid lines in Fig.\;\ref{fig:delsp}) 
agrees well with that given in \citet{adhi14} (dotted lines) in the limit of
small accretion rates. When the accretion rate is large, our results yield larger
$\Delta_{\rm m}^{\rm sp}$ especially at low $\Omega_{\rm m}^{\rm obs}$. 
The difference arises from the different theoretical approaches of the two
methods: while we use the self-similarity ansatz to compute the trajectories of many
shells and the mass profile consistently, \citet{adhi14} imposed a more
realistic NFW profile of the dark matter halo but computed only the trajectory
of one shell without making it consistent with the mass profile. We assess how
much of the difference arises from the different mass profile in
Appendix.\;\ref{app:innerprof}.

\section[]{Summary and Conclusion}
\label{sec:conclusion}
Being dominated by recent accretions rather than by relaxation processes, the
outer profile of dark matter halos can be reasonably well-described by one dimensional
spherical collapse models. The challenge, however, is to consistently treat the
trajectories of dark matter shells and the mass profile of dark matter halos in
a $\Lambda$CDM universe. 

We achieved this by generalising the self-similar spherical collapse model of
\citet{fillmore84} to a $\Lambda$CDM universe. To remedy the loss of strict
self-similarity, we simultaneously computed the trajectories of dark matter
shells that collapsed at different cosmic times using a simple
dimensionless form of the spherical collapse model which reduced the
computational effort.

The resulting dark matter profiles depend on two parameters: the mass accretion
rate of the dark matter halo, and the matter content of the universe (and thus
the redshift).
We find that the shape of dark matter profiles are clearly separated into an inner power
law profile whose slope depends on the mass accretion rate but not on the
redshift, and an accretion region where the profiles depend on both, but can be
described easily by the spherical collapse of a single shell to a good
approximation. These inner and accretion regions are linked by a sharp
transiting region around the splashback radius, where the dark matter density
drops abruptly by a factor of $\sim 4-5$. 

We confirmed, and provided more understanding to previous results on how the
splashback radius depends on the accretion rate and the redshift.  
In particular, the puzzling numerical discovery that the
splashback radius $R_{\rm sp}$ scales well with $R_{\rm 200m}$ rather than
with the more frequently used $R_{\rm 200c}$ is found to be an outcome of the
correlated increase of $\Omega_{\rm m}^{\rm obs}$ and the average halo mass accretion rate
with redshift and their canceling effects on $R_{\rm sp} / R_{\rm 200m}$.

New fitting formulae of the splashback radius in various forms are provided. Our
results can aid the interpretation of observations of diffuse hot gas and weak
gravitational lensing observations in the outskirts of galaxy clusters.
We recommend to use $R_{\rm sp}/R_{\rm 200m}$, which is less sensitive to
the shape of mass profiles, as the quantity to link analytical studies to
simulations and observations. 

\section*{Acknowledgements}
XS thanks Eiichiro Komatsu for carefully reading the manuscript and giving
helpful suggestions, as well as Erwin Lau and Surhud More for discussions and
comments, and Susmita Adhikari for explaining the details of their method.

\bibliographystyle{mn2e}

\appendix

\section[]{Minimalistic dimensionless form of spherical collapse in a
$\Omega_{\rm m}+\Omega_{\rm \Lambda}=1$ universe}
\label{app:spherical_collapse}

We consider the nonlinear evolution of a spherical region with
overdensity $\delta(R_i)$. The evolution of
radius $R$ of the shell encompassing mass $M$ shares the same dynamical equation with the background universe \citep[e.g.][]{mo_book}
\eq{
\frac{\dd^2 R}{\dd t^2} =  - \frac{GM}{R^2} + \frac{\Lambda R}{3}  \,,
\label{eq:d2Rdt2}
}
but has a different energy equation including the effect of its curvature $K$,
\eq{
\br{\frac{\dd R}{\dd t}}^2 = \frac{2 G M_i}{R} + \frac{\Lambda R^2}{3} - K\,.
\label{eq:dRdt2}
}
Here, $\Lambda$ is the cosmological constant, $M_i$ is the initial mass
encompassed by the shell, and $R_i = R(a_i)/a_i$ is the comoving size of
the shell at the initial time when the cosmic scale factor $a=a_i$. Since the
density fluctuations are small in the early universe, $\delta(R_i) \ll 1$, the
mass of the overdense region is $M_i = 4\pi\bar{\rho}_0 R_i^3 / 3$ to a
good approximation, with $\bar{\rho}$ being the mean matter density of the
universe, and subscript `0' indicates today ($a=1$).

\begin{figure}
\centering
    \includegraphics[width=.47\textwidth]{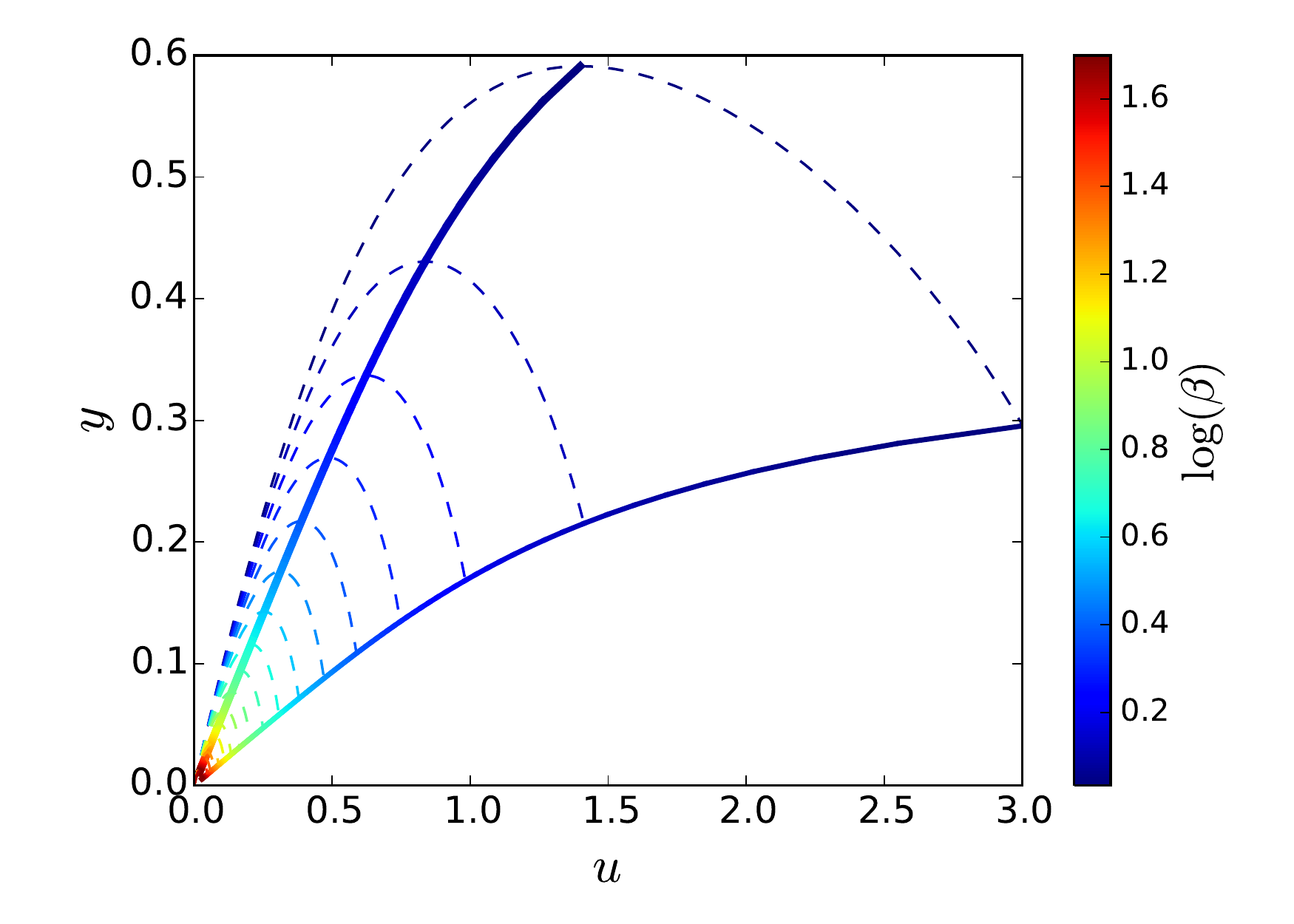} 
  \caption{Spherical collapse trajectories (scaled radius $y$ as a function of
  scaled cosmic scale factor $u$) in an $\Lambda$CDM universe (dashed lines) for
  shells with different scaled initial overdensities $\beta$ (shown with different colors). The turn-around locations of the trajectories are
  marked as the upper solid line, whose analytical expression is given by
  Eqs.\;\ref{eq:yta} and \ref{eq:Iellip}. The locations where the shells
  collapse to half their turn-around radius are shown as the lower
  solid line.}
\label{fig:yu}
\end{figure}

Compared to the background universe with $K=0$, the shell expands more slowly. A
shell encompassing a large overdensity finally turns around at a radius $R_{\rm
ta}$ and starts to infall towards the center (see trajectories of the
shells as dashed lines in Fig.\;\ref{fig:yu}).
Shell-crossing happens after this turn-around and the energy equation
(\ref{eq:dRdt2}) is no longer valid due to the interactions among different
shells.

\begin{table}
 \centering
{
\caption{Summary of the notations for the variables describing the background
cosmology (the first three rows) and the spherical overdense region (the last three rows), both
in their original and scaled forms. In the scaled forms, the dimensional
physical scales defined by the mass of the initial overdense region $M_i$, its comoving
radius $R_i = R(a_i) / a_i$ and the mean overdensity of the universe at
redshift zero $\rho_0$ are taken out of the cosmological constant, time, the
curvature and the radius of the overdense region; and the cosmology as described by the
scaled cosmological constant is taken out of the cosmic scale factor, time,
the overdensity and the radius of the overdense region.}
}
\begin{tabular}{ccc}
 \hline
  & Original & Scaled \\
\hline
Cosmological constant			& $\Lambda$ & $w = \Lambda R_i^3 / (6 G
M_i) = \Omega_{\rm m0}^{-1} -1$ \\
Cosmic scale factor				& $a$ & $u = w^{\frac{1}{3}}a$ \\
Time							& $t$ & $I = (8\pi G \rho_0 w/3)^{\frac{1}{2}} t$ \\
\hline
Curvature  & $K$ & $\kappa = K R_i / (2G M_i)$ \\
Overdensity  & $\delta$ & $\beta = \kappa/\kappa_{\rm min}$ with $\kappa_{\rm
min} = 3w^{\frac{1}{3}}/2^{\frac{2}{3}}$ \\
Radius 							& $R$ & $y = w^{\frac{1}{3}}R/R_i$ \\
\hline
\end{tabular}
\label{tab:notations}
\end{table}

Even before turn-around, the dynamical behavior of the shell is determined by
three parameters: 
\items{
\item the mass / size of the overdense region 
\item its overdensity with respect to the mean matter density, which
determines the curvature of the overdense region when the latter is viewed
as a separate universe \citep[e.g.][]{ichiki12}
\item the value of the cosmological constant which sets a a typical time scale
in the evolution of the universe.}
We shall reduce the number of parameters to gain
general quantitative properties of the spherical collapse.
\citet{eke96} have presented a dimensionless form of the spherical collapse
equations with two parameters: scaled $\Lambda$ and $K$ defined as (see
Table.\;\ref{tab:notations} for a summary of the notations) 
\eq{
 \label{eq:w}
  w := \Lambda R_i^3/ 6GM_i = \Omega_{\rm m0}^{-1} - 1  \to
  \begin{cases}
     \infty  & \quad \text{for } \Omega_{\rm m0} \to 0 \\
     0       & \quad \text{for } \Omega_{\rm m0} \to 1  \text{  (EdS
     universe)}\\
     1       & \quad \text{when } \Omega_{\rm m0} = \Omega_{\rm \Lambda0},\\
  \end{cases}
}

\eq{
\kappa := K R_i/2GM_i \propto \delta \,.
}
They also showed that, in a universe with a non-zero cosmological constant
($w>0$), a minimum initial overdensity which corresponds to $\kappa_{\rm min} =
3w^{1/3} / 2^{2/3}$ is required for a shell to collapse.

We use this to define a new parameter $\beta := \kappa/\kappa_{\rm min}$ to
describe the overdensity of the shell. At the same time, we further scale out
the cosmology (`$w$') dependencies in the radius and time variables:
\eqs{
& y := w^{\frac{1}{3}} R / R_i \,, \\
& I := \br{\frac{8\pi G
\bar{\rho}_0 w}{3}}^{\frac{1}{2}} t  = \br{1-\Omega_{\rm m0}}^{\frac{1}{2}} H_0
t\,,
}
with $H_0$ being the Hubble constant.
The scaled time $I$ has a simple analytical relation with the scaled cosmic
scale factor $u := w^{\frac{1}{3}} a$,
\eq{
I = \frac{2}{3} \arcsinh\br{u^{\frac{3}{2}}} \,.
} 
With these definitions, $u = (1-\Omega_{\rm m})^{\frac{1}{3}}/\Omega_{\rm
m}^{\frac{1}{3}}$ reflects the ratio of dark energy and dark matter at that
time, $y \approx u$ in the early times when the overdensity is small and the overdense region
expand with nearly the same speed as that of the background universe (Fig.\;\ref{fig:yu}).

Written in terms of these scaled parameters, the dynamical
equation (\ref{eq:d2Rdt2}) and the energy equation (\ref{eq:dRdt2}) become 
\eq{
\label{eq:dyn}
\frac{\dd^2 y}{\dd I^2} =
-\frac{M}{2M_i} \frac{1}{y^2} + y \,,
}
and
\eq{
\label{eq:dydI2}
\br{\frac{\dd y}{\dd I}}^2 = 
y^{-1} + y^2 - 3\beta/2^{\frac{2}{3}} = (y-y^*)(y+y^*-1/y/y^*)
}
with $y^*$ being the scaled turn-around radius of the shell \footnote{We will
use superscript `*' to indicate the quantities at the turn-around radius
\textit{of the shell}. In contrast, the quantities at the turn-around
radius \textit{at the time of observation} (i.e. $t^* = t^{\rm obs}$) will be
indicated with subscript `ta'.}. Again, Eq.\;(\ref{eq:dydI2}) no longer
holds after shell-crossing, while Eq.\;(\ref{eq:dyn}) still does as long as the actual
enclosed mass $M(<R)$ is used.

The trajectories of the shells in the absence of shell-crossing until they reach
half of their turn-around radius (`hrta') are shown in Fig.\;\ref{fig:yu}. Now,
a shell is characterised solely by its scaled overdensity $\beta$. The physical scale of the shell is only important in the shell-crossing region. 
Cosmology dependence is removed from the dynamical and
energy equations. Different cosmologies correspond to different values of $u_0$
and $I_0$, or $u_{\rm obs}$ and $I_{\rm obs}$ in a more general sense when we
label a universe with its $\Omega_{\rm m}^{\rm obs}$ value at the time of
observation irrespective of the corresponding redshift. In particular, an EdS
universe corresponds to $u_{\rm obs} \ll 1$ and $I_{\rm obs} \ll 1$, and the
shells collapsed before this time must have $\beta \gg 1$ (Fig.\;\ref{fig:yu}).
A $\Lambda$CDM universe with $\Omega_{\rm m} = 0.3$ has $u_{\rm obs} =
(0.3^{-1}-1)^{1/3} \approx 1.326$.

Due to the simple form of the scaled spherical collapse equation in the
single-stream regime (Eq.\;\ref{eq:dydI2}), useful analytical relations can be
derived among the variables (see Appendix.\;\ref{app:eqs}). These relations also
enable easy determination of initial conditions for the numerical integration of
the dynamical equation in the shell-crossing regime.

\subsection{Analytical relations in the single stream regime of spherical
collapse}
\label{app:eqs}
From the scaled spherical collapse equation Eq.\;(\ref{eq:dydI2}), we can derive
an analytical expression for the scaled turn-around radius $y^*$ as a function
of the scaled overdensity $\beta$, 
\eq{
y^* = 2^{\frac{2}{3}} \beta^{\frac{1}{2}} \sin \br{\frac{1}{3}
\sin^{-1}\br{\beta^{-\frac{3}{2}}}} \,,
\label{eq:yta}
}
or reversely,
\eq{
\beta = \frac{2^{\frac{2}{3}}}{3} \br{y^{*-1} + y^{*2}} \,.
}
For $\beta \ge 1$, $y^*\le 2^{-1/3} $, i.e. there exists a maximum
radius for collapsing regions. In the far future when dark energy completely
dominates the energy content of the universe, all halos will have a truncation
radius beyond which matter can no longer fall onto them (cf. \citealt{sub00}).

Until turn-around, the relation between time and radius is monotonous, which
allows us to express the scaled time as a function of the scaled radius,
\eq{
I = \int_0^{y/y^*} \frac{\sqrt{x}\dd x}{\sqrt{(x-1)\br{x^2 +
x - 1/y^{*3}}}} \,.
}
This form can be written in terms of elliptical integrals. In particular, the
integral to $y=y^*$ can be expressed as the sum of two complete elliptic
integrals $\Pi$ (3rd kind) and \rm{K} (1st kind),
\eq{
\label{eq:Iellip}
I^*  = \frac{1+\sqrt{1+4A}}{\sqrt{A-1/2+\sqrt{1+4A}/2}} \bb{\Pi\br{B, C} -
\rm{K}\br{C} }\,, } 
with $A\equiv1/y^{*3}$, $B\equiv {2}/\br{3+\sqrt{1+4A}}$,
and $C\equiv \sqrt{1+4A}/\br{A-1/2+\sqrt{1+4A}/2}$ (see \citealt{grad65}
3.148.4).

The ratio of the average density within a shell and the mean matter density of
the universe is given by the simple relation 
\eq{
\label{eq:delm}
\Delta_{\rm m} := \frac{\ba{\rho(<y)}}{\bar{\rho}} = \frac{u^3}{y^3}
}
before shell crossing. After shell-crossing, it should be modified to 
\eq{
\Delta_{\rm m} = \frac{u^3}{y^3} \frac{M(<y)}{M_i} \,.
}

In the limit of an EdS universe, i.e. $\beta \to \infty$, we have
\eqs{
y^* &\to 2^{2/3}(3\beta)^{-1} \,,\\
I(y,\beta) &\to y^{*3/2} \bb{\arcsin\sqrt{y/y^*} - \sqrt{(1-y/y^*)y/y^*}} \,,\\
I^* &\to  \pi (3\beta)^{-3/2} \,,\\
y^{*\frac{3}{2}}/I^* &\to 2/\pi \,,\\
\Delta_{\rm m}^* = \frac{u^{*3}}{y^{*3}} &\to \bb{\frac{\sinh^{2/3}\br{{3\pi
(3\beta)^{-3/2}/2}}}{2^{2/3}(3\beta)^{-1}}}^3 \to 9\pi^2/16 \approx 5.55 \,.
}
The last expression on the mean overdensity within the turn-around radius
$\Delta_{\rm m}^*$ reproduces the classical result of \citet{lacey93}. 
When generalized to a $\Lambda$CDM universe, $\Delta_{\rm m}^*$ depends on $\Omega_{\rm m}$, 
\eqs{
\Delta_{\rm m}^{*} & \approx \frac{9\pi^2}{16} + u_{\rm obs}^{2} 
+ 2u_{\rm obs}^{3} \\
& = \frac{9\pi^2}{16} + \br{\frac{1-\Omega_{\rm m}^{\rm
obs}}{\Omega_{\rm m}^{\rm obs}}}^{2/3}  + 2\frac{1-\Omega_{\rm m}^{\rm obs}}{\Omega_{\rm
m}^{\rm obs}} \,, }
with the approximation precise to within 3\% for any value of $u_{\rm obs}$ or
$\Omega_{\rm m}^{\rm obs}$. 

\subsection[]{Dark matter density profile in the single-stream region}
\label{app:single}

With the dynamics of infall described by Eq.\;(\ref{eq:dydI2}), the dark
matter density profile in the single-stream region is fully specified when the
physical scales of the dark matter shells are given, e.g., by the mass
accretion rate of the central halo.
Labeling the dark matter shells by their overdensity parameter $\beta$, the
density at the location of the shell at epoch $u$ is 

\eqs{
\label{eq:accdens}
\rho(u, \beta)  = & \frac{\partial \ln M}{\partial \ln R} \frac{M}{4\pi R^3} 
=  \bar{\rho}(u) \Delta_{\rm m}(u, \beta) \frac{\partial \ln R_i
}{\partial \ln R} \\
= &\bar{\rho}(u) \frac{u^3}{y^3(u, \beta)} \bb{ 1 +
\frac{3}{s}\br{\frac{\dd \ln u_{\rm hrta}(\beta)}{\dd 
\beta}}^{-1}\frac{\partial \ln y}{\partial \beta} }^{-1} \,,
}
whereas its radial location is
\eq{ 
\label{eq:accR}
R(u, \beta) = R_{i} w^{-1/3} y = \br{\frac{3 M_{\rm vir}(u)}{4\pi
\bar{\rho}(u)}}^{\frac{1}{3}} \br{\frac{u_{\rm hrta}(\beta)}{u}}^{\frac{s}{3}}  \frac{y(u, \beta)}{u} \,.
}
They can be easily generalised to other parametrised forms of mass accretion
history.

\section[]{Splashback radius in an EdS universe}
\label{app:splashback_EdS}

As shown by \citet{fillmore84},
in an EdS universe, for a small accretion rate $s \le 3/2$, the gravitational
potential is static; for $s \ge 3/2$, the potential grows, the orbit contracts,
and the amplitude of oscillation decays with time. Namely, since the
turn-around radius and the mass enclosed grow in an EdS universe as $R_{\rm
ta}(I)/R^* = (I/I^*)^{2/3 + 2s/9}$ and $M_{\rm ta}(t)/M^* = (I/I^*)^{2s/3}$
\citep[][Eqs.\;18 and 19]{fillmore84},

\eq{
\frac{M(R, t)}{M^*} = \br{\frac{R}{R^*}}^{\Upsilon}
\br{\frac{R^*}{R_{\rm ta}}}^{\Upsilon} \frac{M_{\rm ta}}{M^*}
= \br{\frac{R}{R^*}}^{\Upsilon} \br{\frac{I}{I^*}}^{\alpha} } with
\[ \alpha = -2\Upsilon/3 -
2\Upsilon s/9 + 2s/3 =
  \begin{cases}
    0  & \quad \text{if } s \le 3/2\\
    4s/9 - 2/3 > 0      & \quad \text{if } s \ge 3/2 \,.\\
  \end{cases}
\]

With these, the dynamical equation describing the oscillation becomes
\eq{
\label{eq:dynhalo}
\frac{\dd^2 R}{\dd I^2} =
- \frac{1}{2w} \br{\frac{I}{I^*}}^{\alpha} 
\frac{R_i^3}{R^{*\Upsilon}} R^{\Upsilon-2} \,.
} 

For time-independent gravitational potentials (when $s \le 3/2$),
Eq.\;(\ref{eq:dynhalo}) does not have explicit time-dependence, and represents
an oscillator with a constant amplitude $R_{\rm sp} = R^*$ and period.
Thus, $I_{\rm sp} \approx 3 I^*$, and the $R_{\rm sp}/R_{\rm ta}$ ratio
reflects only the growth of the turnaround radius with time,  
\eq{
\label{eq:Rsp2ta1}
\frac{R_{\rm sp}}{R_{\rm ta}} =  \frac{R^*}{R_{\rm
ta}} = \br{\frac{I_{\rm sp}}{I^*}}^{-2/3 - 2s/9} \approx 3^{-2/3 - 2s/9}
\,.
}
This relation is precise in the limit of $s\to 0$.

For time-dependent gravitational potentials (when $s > 3/2$), some conserved
quantity is usually exploited to describe the amplitude and period change in the oscillator.
In the limit of slow mass accretion, the radial action of the orbit is
conserved \citep{binney08} even when the gravitational potential grows with
time. More generally, a canonical transformation of coordinates can be
constructed that removes the explicit time-dependence from the Hamiltonian of
the oscillating shell \citep{pena13}. In our case of time varying power-law
potentials with mass profile slope of $\Upsilon=1$ (Eq.\;\ref{eq:gamval}), the
transformation is approximately $R' = ({I}/{I^*})^{\alpha/2} R$ and $\dd \tau =
({I}/{I^*})^{\alpha} \dd I$.
To linear order, the dynamical equation expressed in the transformed quantities 
\eq{
\frac{\dd^2 R'}{\dd \tau^2} = - \frac{1}{2w}  
\frac{R_i^{'3}}{R^{*'}} R^{'-1}
}
represents an oscillator with a period of 
\eq{
\label{eq:tau}
\tau_{\rm p} = 4\sqrt{\pi}\sqrt{w} \frac{R^{*'
3/2}}{R_i^{'3/2}}   = 4\sqrt{\pi} y^{*3/2}\,,
}
and a time-independent amplitude, i.e.
$R'_{\rm sp} = R^{*'}$. Transforming back into the original coordinates, $R_{\rm
sp}/{R^*} = ({I_{\rm sp}}/{I^*})^{-\alpha/2}$. Therefore,

\eq{
\label{eq:Rsp2ta}
\frac{R_{\rm sp}}{R_{\rm ta}} =  \frac{R_{\rm sp}}{R^*} \frac{R^*}{R_{\rm
ta}} = (I_{\rm sp}/I^*)^{-2/3 - 2s/9 - \alpha/2} = (I_{\rm sp}/I^*)^{-\alpha-1}
\,.
}

A shell moving from its turnaround to splashback location completes half an
oscillation, thus
\eqs{
& \tau_{\rm sp} - \tau^* = (I_{\rm sp}^{\alpha+1} - I^{*(\alpha+1)}) /
(\alpha+1) / I^{*\alpha} = \tau_{\rm p}/2  \,,
}
which, together with Eq.\;(\ref{eq:tau}), gives
\eq{
I_{\rm sp}^{\alpha+1} = I^{*\alpha+1} + 2(\alpha+1)\sqrt{\pi}
y^{*3/2} I^{*\alpha} \,.
}
Inserting this into Eq.\;(\ref{eq:Rsp2ta}) finally gives
\eqs{
\label{eq:Rsp2ta2}
\frac{R_{\rm sp}}{R_{\rm ta}}  =  \bb{1
+ 2(\alpha+1)\sqrt{\pi} y^{*3/2} / I^{*}}^{-1} = \bb{1
+ 4(4s/9+1/3)/\sqrt{\pi}}^{-1} \,.
}
Note that this relation is derived under the approximation of a power-law
mass profile which does not capture the full profile shape in the halo
outskirts. In addition, in the limit of extremely high mass accretion
rates, Eq.\;\ref{eq:Rsp2ta2} will fail from neglecting the
non-linear corrections in the coordinate transformations. 

Thus, in summary, we have obtained two analytical approximations for the value
of  ${R_{\rm sp}}/{R_{\rm ta}}$ in an EdS universe as a function of the
accretion rate $s$ (Eqs.\;\ref{eq:Rsp2ta1} and \ref{eq:Rsp2ta2}) suited for $s \le 3/2$ and $s \ge
3/2$ respectively. As shown by Fig.\;\ref{fig:rsp}, they describe the
numerically integrated values rather well for the tested range of $s \le 5$.

\section{Effect of mass profile on the splashback position}
\label{app:innerprof}

\begin{figure}
\centering
  \begin{tabular}{@{}c}
    \includegraphics[width=0.45\textwidth]{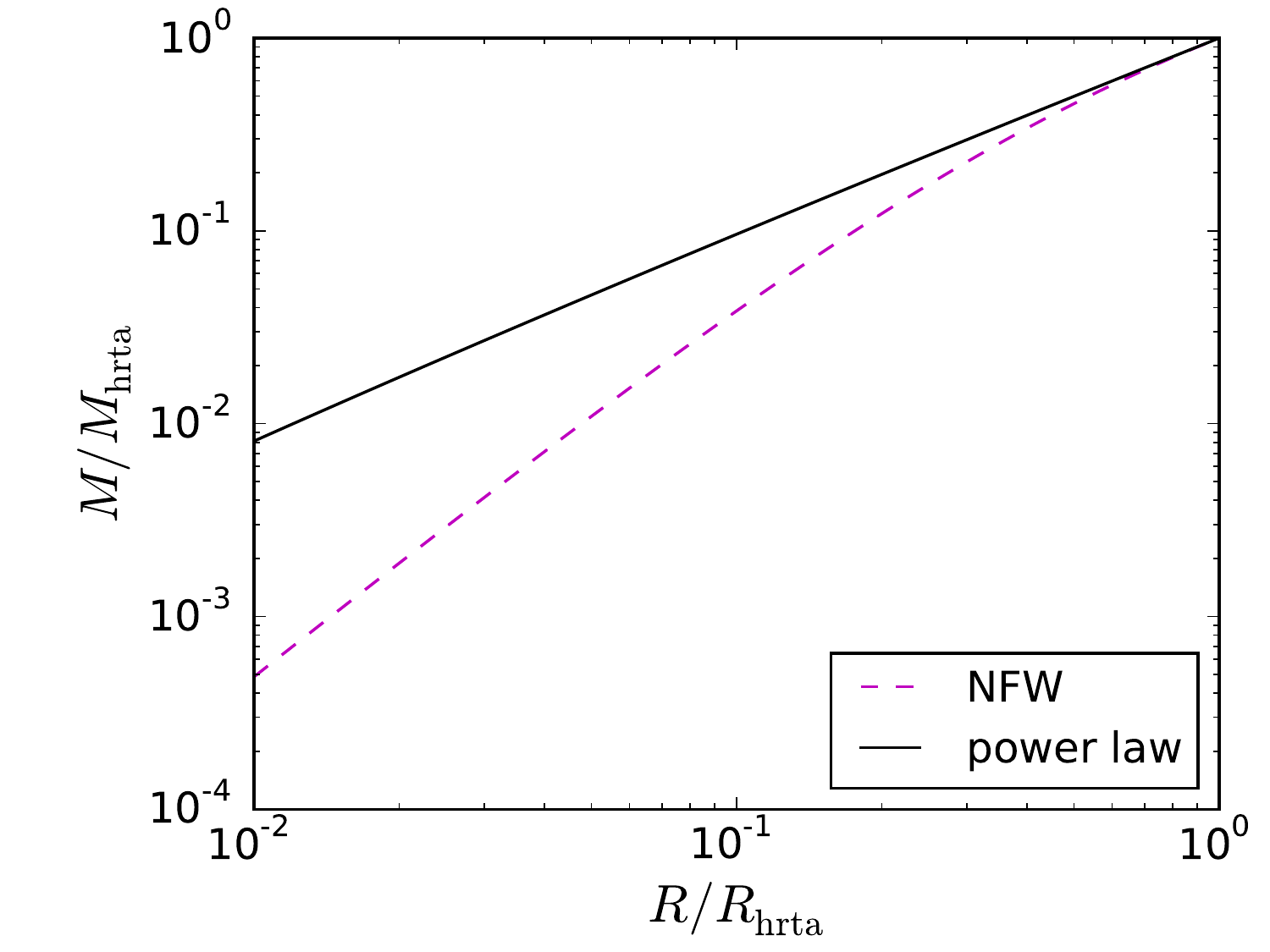}\\   
    \includegraphics[width=0.43\textwidth]{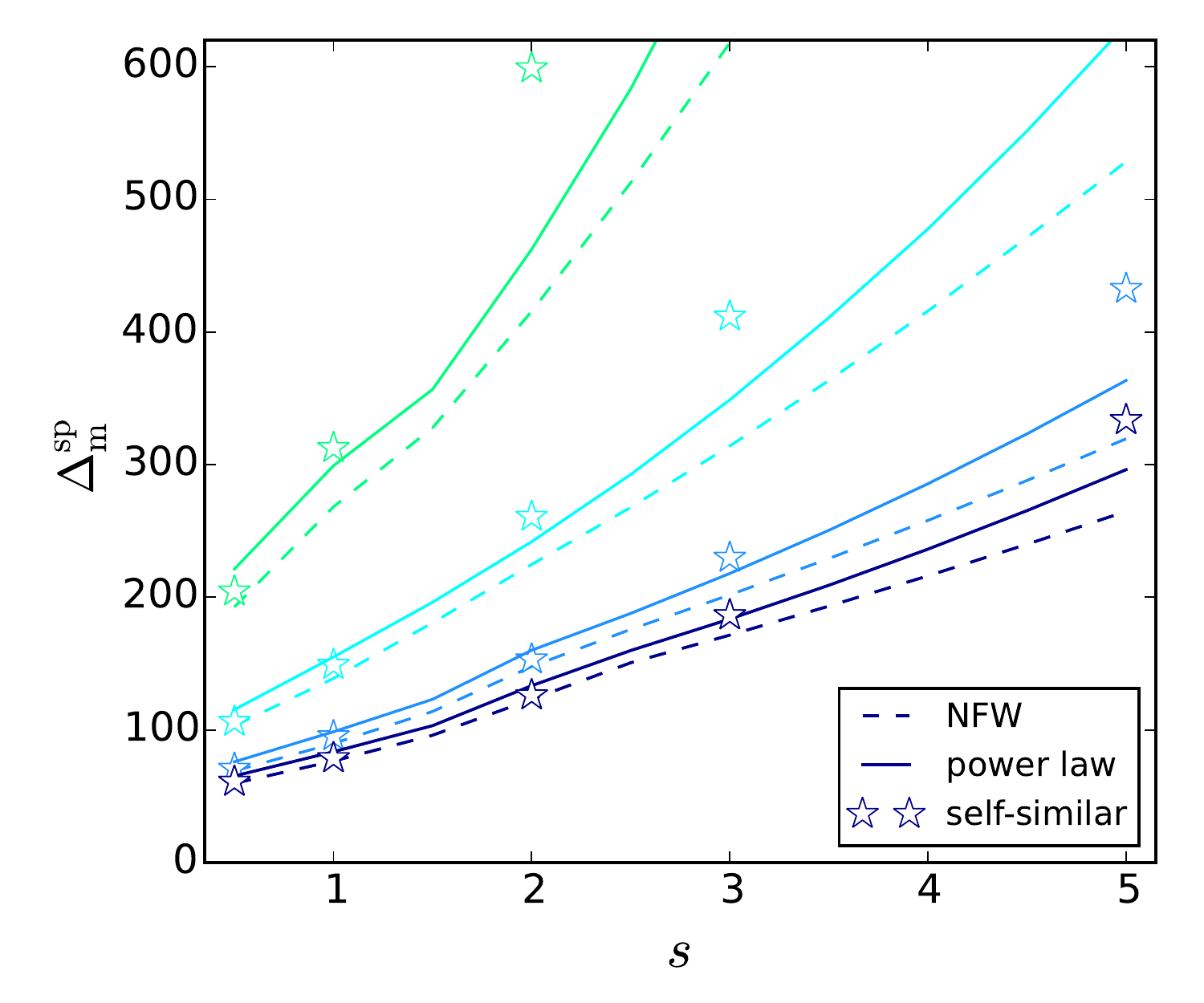}\\
    \includegraphics[width=0.43\textwidth]{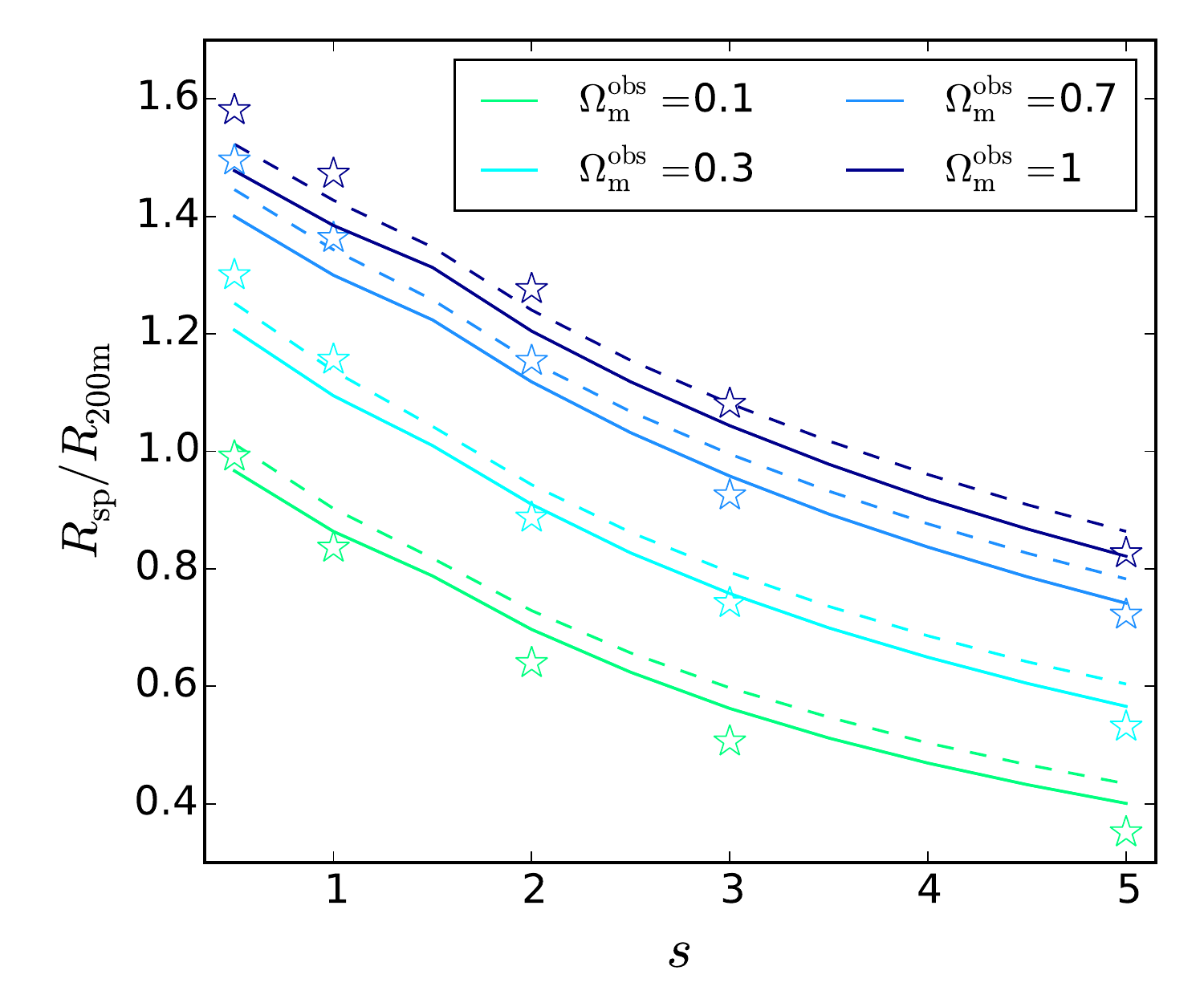} 
  \end{tabular}
  \caption{\textit{Upper panel:} Different mass profiles used for computing the
  splashback radius with the Adhikari et al. method.
  `NFW' presents the profile used in Adhikari et al. paper for mass accretion
  rates $s = 1.5$, and `power law' is the corresponding profile with power-law
  index given by Eq.\;(\ref{eq:gamval}). \textit{Middle panel:} The overdensity
  at the splashback radius. The stars show the results from the self-similar model as
  presented in Fig.\;\ref{fig:delsp}. The dashed lines show reproduced results
  of Adhikari et al., and the solid lines show results using the Adhikari et al.
  method but with power law mass profiles. Different $\Omega_{\rm m}^{\rm obs}$
  values (0.1,0.3,0.7,1) are indicated by different colors, from light green to
  dark blue, respectively.
  \textit{Lower panel:} The splashback radius in terms of $R_{\rm sp}/ R_{\rm 200m}$. Labels are the same as the middle panel.}
\label{fig:sp}
\end{figure}

How do different mass profiles affect the position of splashback? We assess this
by generalising the \citet{adhi14} method: tracing the trajectory of one dark
matter shell through a growing dark matter halo with an NFW profile, to dark
matter halos with different mass profiles (upper panel of Fig.\;\ref{fig:sp}).
The mass profiles considered are constructed to have the same value and
slope at $R_{\rm hrta}$, but different inner slopes. The mass
profile slope at $R_{\rm hrta}$ is derived from the Fillmore \& Goldreich
formula (Eq.\;\ref{eq:gamval}) and thus is similar to that of our self-similar
solutions\footnote{However, this is different from the original \citet{adhi14}
choice, which uses a mass profile slope of $3s/(s+3)$ without the saturation
at $s \ge 3/2$. To differentiate the two, we refer to our reproduced result as
`result from the Adhikari et al. method'. Matching
the mass profile slope of $3s/(s+3)$ without a saturation to an NFW halo would
lead to unrealistically small concentration parameters at relatively large mass
accretion rates, e.g. $c<1$ for $s>2.28$, and significantly smaller
$\Delta_{\rm m}^{\rm sp}$ there.}.

The resulting $\Delta_{\rm m}^{\rm sp}$ as a function of the mass accretion rate
$s$ and the redshift, or more precisely, the matter content of the universe at
the time of observation $\Omega_{\rm m}^{\rm obs}$ is shown as lines with
different line-styles in the middle panel of Fig.\;\ref{fig:sp}.
The difference between them presents the degree of influence of different mass profiles, while
the difference between the solid lines and the stars (our self-similar result)
presents that of the different theoretical assumptions of Adhikari et al. and our
self-similar approach. Both effects matter to similar degree.

Different mass profiles affect $\Delta_{\rm m}^{\rm sp}$ not only through
different gravitational dynamics they lead to, but also simply from the
radius-overdensity conversion. To reduce the second effect, we present the
comparison again in terms of $R_{\rm sp}/R_{\rm 200m}$ (lower panel of
Fig.\;\ref{fig:sp}). In terms of $R_{\rm sp}/R_{\rm 200m}$, the splashback
locations predicted with different mass profiles differ much less. Therefore, we
recommend to use this representation to compare with simulations and
observations. In this representation, the result from the self-similar solution
and that from the Adhikari et al. method agree to better than 10\% for all mass
accretion rates examined when $\Omega_{\rm m}^{\rm obs} \ge 0.3$.

\end{document}